# New analytic criterion for porous solids with pressure-insensitive matrix

Oana Cazacu[*] and Benoit Revil-Baudard

Department of Mechanical and Aerospace Engineering, University of Florida, REEF,

1350 N. Poquito Rd, Shalimar, FL 32579, USA.

## Abstract

In this paper, we address the question of how the relative weighting of the two invariants of the plastic deformation of the matrix influence the mechanical response of a porous metallic material. To this end, we first propose a new isotropic potential for description of the plastic behavior of the matrix that depends on both invariants of the strain-rate deviator. The relative weight of the two invariants is described by a material parameter β. Depending on the sign of the parameter β, the new plastic potential for the matrix is either interior to von Mises strain-rate potential (β <0), coincides with it (β=0) or it is exterior to it. Next, an analytic criterion for a porous solid with matrix governed by the new strain-rate potential is obtained using rigorous upscaling methods. Analysis is conducted for both tensile and compressive axisymmetric loading scenarios and spherical void geometry. No simplifying approximations are considered when estimating the local and overall plastic dissipation, respectively. It is shown that the value of β has a drastic influence on all aspects of the mechanical response. There is a value $\beta = \beta_* < 0$ such that there is almost no influence of $J_3^\Sigma$ on the mechanical response of the porous solid. If the matrix is characterized by $\beta > \beta_*$, the response of the porous material for tensile loadings and $J_3^\Sigma \geq 0$ is softer than that for loadings at $J_3^\Sigma \leq 0$. The reverse holds true for $\beta < \beta_*$. The noteworthy result is that irrespective of the value of the parameter β, the response of the porous solid is harder than that of a porous Tresca material. However, depending on the value of β the rate of void growth or collapse can be either faster or slower than that of a porous Mises material.

[*] Corresponding author: Tel: +1 850 833 9350; fax: +1 850 833 9366.

E-mail address: cazacu@reef.ufl.edu

## 1. INTRODUCTION

Based on micromechanical considerations, Gurson (1977) have demonstrated that the presence of voids results in plastic deformation being accompanied by volume changes. Furthermore, using rigorous limit analysis theorems, Gurson derived a stress potential for porous solids with von Mises matrix (Gurson, 1975; Gurson, 1977). This potential involves only dependence of the mean stress and of the second-invariant of the stress deviator. Various modifications of the original Gurson (1977) model have been proposed such as to account for the dependence of yielding on the third-invariant of the stress deviator, $J_3^{\Sigma}$ (e.g. Nashon and Hutchinson, 2008; Nielsen and Tvergaard, 2009). Very recently, Cazacu et al. (2013) demonstrated that the exact yield criterion of porous material with von Mises matrix and randomly distributed spherical voids ought to be centro-symmetric, and should involve a very specific coupling between the mean stress and the third-invariant of the stress deviator. Using rigorous limit-analysis theorems, analytic yield criteria that capture the aforementioned features of the yielding behavior were developed (for axisymmetric loadings, see Cazacu et al. 2013, for full 3-D loadings, see Revil-Baudard and Cazacu, 2014(a), Cazacu and Revil-Baudard, 2015). Because these yield criteria for a porous von Mises material involve coupling between the mean stress and shear stresses, the rate of void growth and collapse is influenced by $J_3^{\Sigma}$. Nevertheless, the sensitivity to $J_3^{\Sigma}$ is rather small. For a porous Tresca material, very recently, Cazacu et al. (2014), Revil-Baudard and Cazacu (2014) have derived analytic plastic potentials expressed in the stress and strain-rate space, respectively. It was shown that the response of a porous Tresca material is much softer than that of a porous Mises material. The rate of void growth in a material with a Tresca matrix can be 20% faster than the rate of void growth in a porous Mises material (see also Revil-Baudard and Cazacu, 2014 (b)).

It is well known that the plastic behavior of any isotropic fully-dense material is described by a combination of the second and third-invariant of the plastic deformation (Prager, 1945). The fundamental question that we pose and address in this paper is: how the relative "weighting" of the two invariants of the plastic deformation of the matrix affect the mechanical response of the porous solid. Specifically, how strong is the influence of the third-invariant of the stress deviator, $J_3^{\Sigma}$, on the overall response of the porous solid. Most importantly, how does the rate of void evolution compare to that in a porous Mises or a porous Tresca material.

The structure of the paper is as follows. First, we propose a potential for description of the plastic behavior of the matrix that depends on both invariants of the strain-rate deviator (Section 2.2). This new plastic potential involves a unique parameter β that provides proper weighting between these invariants. While Tresca's strain-rate potential is always an upper bound, depending on the sign of the parameter β, this new strain-rate potential is either interior to the von Mises strain-rate potential (β <0), coincides with it (β = 0) or is exterior to it (β > 0). Next, using rigorous limit-analysis theorems (briefly recalled in Section 2.1), we derive a new criterion for a porous solid with matrix governed by the new strain-rate plastic potential introduced in Section 2.2. It is shown that although the criterion for the matrix is very general and involves dependence on both invariants it is possible to derive a closed-form expression for the criterion for the porous solid (Section 3). Analysis of the mechanical response of the porous solid according to the new criterion is examined both in terms of yielding and void growth (Section 4). It is shown that the sensitivity to the third-invariant of the plastic deformation of the matrix, described by the parameter β, plays a paramount importance. The noteworthy result is that the softest response is that of a porous Tresca material (Section 5). However, depending on the value of β the rate of

void growth or collapse can be either faster or slower than that of a porous Mises material. The main findings of this paper are summarized in Section 6.

Regarding notations, vector and tensors are denoted by boldface characters. If **A** and **B** are second-order tensors, the contracted tensor product between such tensors is defined as: $\mathbf{A}:\mathbf{B} = \mathrm{A_{ij}B_{ij}}$ i, j = 1…3, tr denotes the trace of the tensor.

## 2. Statement of the problem

### 2.1. Kinematic homogenization framework for development of plastic potentials for porous metallic materials

A general framework for deriving the plastic potential of a porous solid with matrix described by any convex plastic potential is that of kinematic limit analysis. The main theorems that will be further used for the derivation of the new criterion for porous solids are briefly recalled in the following.

Consider a representative volume element $\Omega$, composed of a homogeneous rigid-plastic matrix material that is incompressible and a traction-free void. The matrix material is described by a convex potential $\psi(\mathbf{d})$ in the strain-rate space, homogeneous of degree one with respect to positive multipliers, such that

$$\boldsymbol{\sigma} = \nu \frac{\partial \psi}{\partial \mathbf{d}} . \qquad (1)$$

In Eq. (1) $\boldsymbol{\sigma}$ is the Cauchy stress tensor, $\mathbf{d} = \frac{1}{2}\left(\nabla \mathbf{v} + \nabla \mathbf{v}^T\right)$ denotes the strain rate tensor, and $\mathbf{v}$ is the velocity field. As shown by Ziegler (1977) and Hill (1987), these equations express the fact that $\psi(\mathbf{d})$ is indeed equal to the plastic dissipation potential,

$$\pi(\mathbf{d}) = \sup_{\boldsymbol{\sigma} \in C}\left(\sigma_{ij} d_{ij}\right) \quad . \qquad (2)$$

with i, j = 1…3, C being the convex domain delimited by the yield surface, and sup stands for supremum.

Let $f$ be the void volume fraction. In the present study, we assume uniform strain rate boundary conditions on $\partial\Omega$, i.e.

$$\mathbf{v} = \mathbf{D}\cdot\mathbf{x}, \text{ for any } \mathbf{x}\in\partial\Omega \tag{3}$$

with **D** constant. For the boundary conditions (3), the Hill-Mandel (Hill, 1967; Mandel, 1972) lemma applies; hence,

$$\left\langle \sigma_{ij} d_{ij} \right\rangle_{\Omega} = \Sigma_{ij} D_{ij}, \tag{4}$$

where $\langle\ \rangle$ denotes the average value over the representative volume $\Omega$, and $\boldsymbol{\Sigma} = \langle \boldsymbol{\sigma} \rangle_{\Omega}$. Furthermore, it has been shown (see Talbot and Willis, 1985) that there exists a macroscopic strain rate potential $\Pi = \Pi(\mathbf{D})$ such that

$$\boldsymbol{\Sigma} = \frac{\partial \Pi(\mathbf{D})}{\partial \mathbf{D}} \text{ with } \Pi(\mathbf{D}) = \inf_{\mathbf{d}\in K(\mathbf{D})} \left\langle \pi(\mathbf{d}) \right\rangle_{\Omega} \quad . \tag{5}$$

In Eq. (5), inf stands for infimum, the minimization being done over $K(\mathbf{D})$, which is the set of incompressible velocity fields compatible with homogeneous strain-rate boundary conditions, i.e. condition (3) applies. These theorems will be further used for the derivation of the plastic potential of a porous solid containing a random distribution of spherical voids. The plastic behavior of the matrix material will be described by a new strain-rate potential that depends on both invariants of the local plastic strain-rate tensor, **d**.

### 2.2. New isotropic strain rate potential for isotropic pressure-insensitive materials

Plastic potentials expressed in terms of strain rates have been shown to be very versatile for the description of the plastic deformation. Concerning the use of such potentials in the context of multi-scale crystal plasticity the reader is referred to the work of van Houtte and collaborators, e.g. Van Houtte, 1994; for examples of strain-rate potentials developed in the framework of the mathematical theory of plasticity and applied to metal forming see Hill (1987), Barlat et al., 1993; Chung et al., 1997; Cazacu et al., 2010, etc.).

In this paper, it is proposed a new plastic strain-rate potential for isotropic fully-dense materials ( tr(**d**) = 0). The expression of this strain-rate potential is:

$$\psi(\mathbf{d}) = \frac{\sqrt{j_2}}{B}\left(1+\beta\frac{j_3^2}{j_2^3}\right), \tag{6}$$

with

$$B = \frac{1+4\beta/27}{\sqrt{4/3}}. \tag{7}$$

In Eq. (6), $\mathrm{j}_2 = \frac{1}{2}\mathbf{d}{:}\mathbf{d}$ is the second-invariant of the plastic strain-rate tensor **d**, $\mathrm{j}_3 = \det(\mathbf{d})$ is the third-invariant of **d**, whereas β is a model parameter. The constant *B* appearing in the expression of the criterion depends solely on β and is defined such that for uniaxial tension $\psi(\mathbf{d})$ is equal to the axial strain rate. It is worth noting that for β=0, the proposed strain-rate potential given by Eq.(6) reduces to the strain-rate potential associated to the von Mises stress potential, $\varphi_{\mathrm{Mises}}(\boldsymbol{\sigma}) = \sqrt{(3/2)\boldsymbol{\sigma}'{:}\boldsymbol{\sigma}'}$. Indeed, for β=0 :

$$\psi(\mathbf{d}) = \sqrt{\frac{4}{3}\mathrm{j}_2} = \sqrt{(2/3)\mathbf{d}{:}\mathbf{d}} = \dot{\varepsilon}, \tag{8}$$

where $\dot{\varepsilon}$ denotes the von Mises equivalent strain-rate and $\boldsymbol{\sigma}'$ the stress deviator. If $\beta \neq 0$, the new plastic potential depends on both invariants of **d**. For $\psi(\mathbf{d})$ to be convex, the range of variation of β is:

$$\frac{-9}{24} \le \beta \le \frac{27}{68} . \qquad (9)$$

Let $(\mathbf{e}_1,\mathbf{e}_2,\mathbf{e}_3)$ be a Cartesian coordinate system associated with the principal directions of **d**. To represent the cross-section of $\psi(\mathbf{d})$ with the octahedral plane (i.e. the plane of normal **n** = $\frac{1}{\sqrt{3}}\mathbf{e_1}+\frac{1}{\sqrt{3}}\mathbf{e_2}+\frac{1}{\sqrt{3}}\mathbf{e_3}$), it is convenient to introduce the Oxyz frame of unit vectors ($\mathbf{e}_x$, $\mathbf{e}_y$, $\mathbf{e}_z$), which are related to the eigenvectors ($\mathbf{e}_1$, $\mathbf{e}_2$, $\mathbf{e}_3$) by the following relations:

$$\mathbf{e}_x = \frac{1}{\sqrt{3}}(\mathbf{e}_1+\mathbf{e}_2+\mathbf{e}_3),\ \ \mathbf{e}_y = -\frac{1}{\sqrt{2}}(\mathbf{e}_1 - \mathbf{e}_2),\ \ \mathbf{e}_z = \frac{1}{\sqrt{6}}(2\mathbf{e}_3-\mathbf{e}_1-\mathbf{e}_2). \qquad (10)$$

Any arbitrary state on the isosurface $\psi(\mathbf{d})$=constant, say $P(d_1,d_2,d_3)$, is characterized by two polar-type coordinates, $(R,\gamma)$, where

$$R = |OP| = \sqrt{d_1^2+d_2^2+d_3^2} = \sqrt{2j_2}\ , \qquad (11a)$$

while $\gamma$ denotes the angle between $\mathbf{e}_y$ and **OP**, so

$$\tan(\gamma) = \sqrt{3}\,\frac{d_3}{d_2 - d_1}\ , \qquad (11b)$$

with ($d_1$, $d_2$, $d_3$) being the principal values of **d**. Let $\mathbf{f}_i$ be the projections of the eigenvectors $\mathbf{e}_i$, i = 1…3 on the octahedral plane. Obviously, $\mathbf{f}_3$ = $\mathbf{e}_z$ (see Eq. (10) and Fig.1). Because $\psi(\mathbf{d})$ is isotropic and even function of **d** (see Eq.(6)), the projection of the potential in the octahedral plane has six-fold symmetry. Therefore, it is sufficient to determine the shape of the cross-section of the surface $\psi(\mathbf{d})$=constant, i.e. $R = R(\gamma)$, only in the sector $-\pi/6 \le \gamma \le \pi/6$. The shape in all the other sectors is then obtained by symmetry arguments. Note that on the surface axisymmetric states correspond to either $\gamma = -\pi/6$ $(d_1 = d_3 < d_2)$ or $\gamma = \pi/6$ $(d_2 = d_3 > d_1)$ while shear loading ($d_3$=0) corresponds to $\gamma$ =0.

As an example, in Fig. 1 are shown the representation in the octahedral plane of the potential given by Eq.(6) for several values of the parameter $\beta = -0.35, -0.15, 0$ (von Mises), 0.2, and 0.38, respectively . Note that for $\beta \neq 0$, the cross-sections are hexagons with rounded corners.

As concerns the limiting values of the parameter $\beta$, i is worth noting that if $\beta$ = 27/68, the curvature of the cross-section is zero for axisymmetric states; if $\beta$ = -9/24, the curvature is zero for shear loading ( $\gamma$ =0).

It is also worth comparing the new strain-rate potential with the strain-rate potential associated with Tresca's maximum shear stress criterion, i.e.

$$\psi_{\text{Tresca}}(\mathbf{d}) = \left(|d_1| + |d_2| + |d_3|\right)/2\,, \tag{12}$$

(for the derivation of Eq. (12), see for example, Lubliner (2008)). While the projection in the octahedral plane of the Mises strain rate potential (Eq.(8)) is a circle, the projection of Tresca's strain rate potential is a regular hexagon (see (Eq. (12)) with the Tresca's hexagon circumscribing the von Mises circle. As an example, in Fig. 2 are represented in the octahedral plane the projection of the proposed strain-rate potential corresponding to several values of the parameter $\beta$, along with the von Mises ($\beta$=0) and Tresca strain rate potential, respectively. It is to be noted that irrespective of the value of the parameter $\beta$, the Tresca strain-rate potential is an upper bound (exterior to the other surfaces). For $\beta$>0, the proposed strain-rate potential lies between the von Mises and the Tresca potential, its shape evolving from a circle for $\beta$=0 to a hexagon with rounded corners for $\beta$ >0 (see for example, Fig.2(a)). It is very interesting to note that for $\beta$ <0, the von Mises strain-rate potential is exterior to the new strain rate potential, the smaller the value of $\beta$, the stronger the deviation of the surface from the von Mises circle (e.g. see Fig. 2(c) for $\beta$=–0.15 and Fig. 2(d) for $\beta$=–0.35).

**3. Derivation of the analytic yield criterion for porous aggregate with matrix described by a plastic potential depending on both invariants**

Using the kinematic homogenization approach presented in Section 2.1., we will now derive in closed-form a plastic potential for isotropic materials containing randomly distributed spherical voids for which the matrix plastic behavior is described by the new strain-rate potential $\psi(\mathbf{d})$ given by Eq. (6). It is worth noting that since $\psi(\mathbf{d})$ is an even function of the local strain rate tensor **d**, it follows that the exact macroscopic strain-rate potential of the porous solid, $\Pi = \Pi(\mathbf{D}, f)$ would also be an even function of the macroscopic strain rate tensor **D**, and that the macroscopic yield function of the porous aggregate, which is defined as

$$F(\mathbf{\Sigma}, f) = \sup_{\mathbf{D}} \left[ \mathbf{\Sigma}:\mathbf{D} - \Pi(\mathbf{D}, f) \right] \tag{13}$$

is also an even function. Indeed,

$$F(-\mathbf{\Sigma}, f) = \sup_{\mathbf{D}} \left[ -\mathbf{\Sigma}:\mathbf{D} - \Pi(\mathbf{D}, f) \right] = \sup_{\mathbf{D}} \left[ \mathbf{\Sigma}:(-\mathbf{D}) - \Pi(-\mathbf{D}, f) \right] = F(\mathbf{\Sigma}, f) \tag{14}$$

Because the voids are spherical and randomly distributed in the matrix, the exact macroscopic yield function of the porous solid, $F(\mathbf{\Sigma}, f)$, ought to be isotropic. By the usual arguments based on theorems of representation of scalar isotropic functions (e.g. Boehler, 1987), it follows that $F(\mathbf{\Sigma}, f)$ should depend on the stress tensor $\mathbf{\Sigma}$ only through its invariants, i.e. the mean stress, the second and the third-invariant of the stress deviator, respectively, i.e.

$$F(\mathbf{\Sigma}, f) = F\left(\Sigma_m, \Sigma_e, J_3^{\Sigma}, f\right),$$

where $\Sigma_m = \frac{1}{3} tr(\mathbf{\Sigma})$ , $\Sigma_e = \sqrt{3 J_2^\Sigma}$ and $J_3^\Sigma = \frac{1}{3} tr(\Sigma')^3$, with $\Sigma'$ being the stress deviator. Since $F(\mathbf{\Sigma}, f)$ is an even function, it follows that

$$F(\Sigma_m, \Sigma_e, J_3^\Sigma, f) = F(-\Sigma_m, \Sigma_e, -J_3^\Sigma, f) \qquad (15)$$

which means that the yield surface of the porous solid is centro-symmetric. Furthermore, only for purely hydrostatic loading ($\mathbf{\Sigma}' = 0$), the yield locus is symmetric with respect to the axis $\Sigma_m = 0$.

As mentioned, we assume that the porous solid contains randomly distributed spherical voids, hence a representative volume element (RVE) is a hollow sphere of inner radius, *a*, and outer radius, *b*. Thus, the void volume fraction $f = (a/b)^3$. The limit analysis will be conducted for both tensile and compressive states, the overall strain rate **D** being considered to be axisymmetric. We use the trial velocity field **v**, deduced by Rice and Tracey (1969), namely

$$\mathbf{v} = \frac{b^3}{r^2} D_m \mathbf{e}_r + \mathbf{D}'\mathbf{x}, \qquad (16)$$

where **x** is the Cartesian position vector that denotes the current position in the sphere, $\mathbf{e}_r$ is the radial unit vector, $r = \sqrt{x_1^2 + x_2^2 + x_3^2}$ is the radial coordinate; $D_m = \frac{1}{3}(2D_{11} + D_{33})$, while $\mathbf{D}'$ is the deviator of **D**. Note that this velocity field is isotropic, satisfies uniform strain-rate boundary conditions (Eq. (3)), and is incompressible, i.e.

$$\mathbf{v}(\mathbf{x} = b\mathbf{e}_r) = \mathbf{D}\mathbf{x} \text{ and } \operatorname{div}(\mathbf{v}) = 0. \qquad (17)$$

Let us denote by $\Pi^+(\mathbf{D}, f)$ the macroscopic plastic dissipation corresponding to Rice and Tracey's velocity field **v** given by Eq. (16) i.e.

$$\Pi^+(\mathbf{D}, f) = \frac{1}{V}\int_\Omega \pi(\mathbf{d})\, d\Omega = \frac{\sigma_0}{V}\int_\Omega \psi(\mathbf{d}) = \frac{\sigma_0}{BV}\int_\Omega \sqrt{j_2}\left(1 + \beta \frac{j_3^2}{j_2^3}\right) d\Omega, \qquad (18)$$

where $\sigma_0$ is the yield value in uniaxial tension, $V = 4\pi b^3/3$ is the volume of the RVE, and the expression of the parameter $B$ in terms of β is given by Eq.(7). The macroscopic stresses associated with $\Pi^+(\mathbf{D},f)$ are then given by:

$$\Sigma_{11} = \Sigma_{22} = \frac{\partial \Pi^+}{\partial D_{11}} \quad \text{and} \quad \Sigma_{33} = \frac{\partial \Pi^+}{\partial D_{33}}$$

or

$$\Sigma_e = \left|\Sigma_{11} - \Sigma_{33}\right| = \left|\frac{\partial \Pi^+}{\partial D_e}\right| \quad \text{and} \quad \Sigma_m = 2\Sigma_{11} + \Sigma_{33} = \frac{1}{3}\left|\frac{\partial \Pi^+}{\partial D_m}\right| \tag{19}$$

where $D_e = \sqrt{2D'_{ij}D'_{ij}/3} = 2\left|D'_{11}\right|$ is the macroscopic equivalent strain rate.

Eq. (19) constitute the expressions of the new criterion for the porous solid.

Note that obtaining a closed-form expression of the criterion hinges on the analytical calculation of the integral representing $\Pi^+(\mathbf{D},f)$ (see Eq. (18)-(19)). The major difficulty in evaluating this integral is due to the fact that for $\beta \neq 0$, the local plastic dissipation depends on both invariants of the local strain-rate tensor, **d** (see Eq. (6)) and thus the calculations are much more complicated than in the case when the matrix plastic behavior is governed by the von Mises criterion ($\beta$=0).Hence, the principal values (unordered) of the local strain-rate field $\mathbf{d} = \frac{1}{2}\left(\nabla\mathbf{v} + \nabla\mathbf{v}^T\right)$ corresponding to the velocity field $\mathbf{v}$ given by Eq. (16) are:

$$\begin{cases} d_1 = D'_{11} + D_m\left(\frac{b}{r}\right)^3 \\ d_2 = -\frac{1}{2}\left(D'_{11} + D_m\left(\frac{b}{r}\right)^3\right) + \frac{3}{2}\sqrt{D'^2_{11} + D^2_m\left(\frac{b}{r}\right)^6 + 2D'_{11}D_m\left(\frac{b}{r}\right)^3\cos 2\theta} \quad , a \le r \le b \\ d_3 = -\frac{1}{2}\left(D'_{11} + D_m\left(\frac{b}{r}\right)^3\right) - \frac{3}{2}\sqrt{D'^2_{11} + D^2_m\left(\frac{b}{r}\right)^6 + 2D'_{11}D_m\left(\frac{b}{r}\right)^3\cos 2\theta} \end{cases}$$

(20)

Next, calculation of $j_2 = \frac{1}{2}\left(d_1^2 + d_2^2 + d_3^2\right)$ and $j_3 = d_1 d_2 d_3$, and substitution in the expression of the matrix plastic potential (Eq. (6)), leads to the following expression of the plastic dissipation associated to the trial velocity field $\mathbf{v}$:

$$\pi(\mathbf{d}) = \frac{\sqrt{3}\sigma_0}{B}\left(D_m^{\ 2}\left(\frac{b}{r}\right)^6 + D'_{11}D_m\left(\frac{b}{r}\right)^3\left(3\cos(\theta)^2 - 1\right) + D'^{\ 2}_{11}\right)^{1/2} \left(1 + \frac{\beta}{27}\frac{\left(D_m\left(\frac{b}{r}\right)^3 + D'_{11}\right)^2\left(2D_m^{\ 2}\left(\frac{b}{r}\right)^6 + D_m D'_{11}\left(\frac{b}{r}\right)^3\left(9\cos(\theta)^2 - 5\right) + 2D'^{\ 2}_{11}\right)^2}{\left(D_m^{\ 2}\left(\frac{b}{r}\right)^6 + D_m D'_{11}\left(\frac{b}{r}\right)^3\left(3\cos(\theta)^2 - 1\right) + D'^{\ 2}_{11}\right)^3}\right)$$

(21)

It is worth noting that given the expression of $\pi(\mathbf{d})$ it follows that $\Pi^+(\mathbf{D}, f)$ given by Eq. (18) is an even function, invariant under the transformation: $\left(D_m, D'_{11}\right) \to \left(-D_m, -D'_{11}\right)$. Thus, the potential of the porous material, $\Pi^+(\mathbf{D}, f)$, needs to be estimated only for the cases when ($D_m \ge 0$, $D'_{11} > 0$) and ($D_m \ge 0$, $D'_{11} < 0$), the stresses at yielding corresponding to all the other strain paths being subsequently obtained by symmetry. Moreover, the resulting yield criterion for the porous solid (19) associated to $\Pi^{+}(\mathbf{D}, f)$ has all the remarkable symmetry properties (15) of the exact yield criterion, which is associated to the exact potential $\Pi(\mathbf{D},f)$ (see also theorem given by Eq.(5), Section 2.1).

For ($D'_{11} > 0$ and $D_m \geq 0$), the local plastic dissipation given by Eq. (21) writes:

$$\pi(\mathbf{d})=\frac{\sigma_0\sqrt{3}}{2B}\left(4D_m{}^2\left(\frac{b}{r}\right)^6+2D_eD_m\left(\frac{b}{r}\right)^3\left(3\cos(\theta)^2-1\right)+D_e{}^2\right)^{1/2}$$
$$\left(1+\frac{\beta}{27}\frac{\left(2D_m\left(\frac{b}{r}\right)^3+D_e\right)^2\left(4D_m{}^2\left(\frac{b}{r}\right)^6+2D_mD_e\left(\frac{b}{r}\right)^3\left(9\cos(\theta)^2-5\right)+D_e{}^2\right)^2}{\left(4D_m{}^2\left(\frac{b}{r}\right)^6+2D_mD_e\left(\frac{b}{r}\right)^3\left(3\cos(\theta)^2-1\right)+D_e{}^2\right)^{5/2}}\right) \tag{22}$$

It follows that $\frac{\partial\Pi^+}{\partial D_{11}} > \frac{\partial\Pi^+}{\partial D_{33}}$ and $\frac{\partial\Pi^+}{\partial D_m} > 0$ . Next, using Eq. (19), we obtain that the stresses at yielding of the porous solid are such that $\Sigma_{11} > \Sigma_{33}$, i.e. the third-invariant of the stress deviator $J_3^\Sigma = -\frac{2}{27}(\Sigma_{11}-\Sigma_{33})^3 < 0$, and the mean stress $\Sigma_m > 0$.

On the other hand, if ($D'_{11} < 0$ and $D_m \geq 0$) the local plastic dissipation given by Eq. (21) writes:

$$\pi(\mathbf{d})=\frac{\sigma_0\sqrt{3}}{2B}\left(4D_m{}^2\left(\frac{b}{r}\right)^6-2D_eD_m\left(\frac{b}{r}\right)^3\left(3\cos(\theta)^2-1\right)+D_e{}^2\right)^{1/2}$$
$$\left(1+\frac{\beta}{27}\frac{\left(2D_m\left(\frac{b}{r}\right)^3-D_e\right)^2\left(4D_m{}^2\left(\frac{b}{r}\right)^6-2D_mD_e\left(\frac{b}{r}\right)^3\left(9\cos(\theta)^2-5\right)+D_e{}^2\right)^2}{\left(4D_m{}^2\left(\frac{b}{r}\right)^6-2D_mD_e\left(\frac{b}{r}\right)^3\left(3\cos(\theta)^2-1\right)+D_e{}^2\right)^{5/2}}\right) \tag{23}$$

so $\frac{\partial \Pi^+}{\partial D_{33}} < \frac{\partial \Pi^+}{\partial D_{11}}$ and $\frac{\partial \Pi^+}{\partial D_m} > 0$ which corresponds to $\Sigma_m > 0$ and $J_3^{\Sigma} > 0$. In the following, we will estimate $\Pi^+(\mathbf{D},f)$, and obtain the expression of the yield surface of the porous solid.

Let denote by

$$u = \frac{2|D_m|}{D_e}, \tag{24}$$

the absolute value of the strain rate-triaxiality.

(i) For stress states such that $\Sigma_m \geq 0$ and $J_3^{\Sigma} \leq 0$:

further substitution of Eq. (22) into Eq. (18) and the change of variable

$$y = u\left(\frac{b}{r}\right)^3, \ \alpha = \cos\theta, \tag{25}$$

leads to the following expression of the overall plastic dissipation,

$$\Pi^+(\mathbf{D}, f) = \frac{\sqrt{3}}{4B}(uD_e)\int_u^{u/f}\int_{-1}^{1}\left(\sqrt{y^2+(3\alpha^2-1)y+1} + \frac{\beta}{27}\frac{(y+1)^2\left(2y^2+(9\alpha^2-5)y+2\right)^2}{\left(y^2+(3\alpha^2-1)y+1\right)^{5/2}}\right)\frac{dy}{y^2}d\alpha$$

This integral can be calculated analytically, i.e.

$$\Pi^+(\mathbf{D}, f) = \frac{\sqrt{3}uDe}{4B}\left(F_1\left(\frac{u}{f}\right) - F_1(u)\right) \tag{26}$$

with

$$\begin{aligned} F_1(y) = &-\frac{2\sqrt{3}}{3}\beta \arctan\left(\frac{2y-1}{\sqrt{3}}\right) + \frac{2\sqrt{3}}{9}\left(1+\frac{11}{3}\beta\right)\left(\arctan\left(\sqrt{3}+2\sqrt{y}\right) - \arctan\left(-\sqrt{3}+2\sqrt{y}\right)\right) + \\ &\left(\frac{9y^{3/2}+\sqrt{3}-3y\sqrt{3}-3y^2\sqrt{3}}{9y^{3/2}} - \beta\frac{6y^2\sqrt{3}+9y^{3/2}-12y\sqrt{3}-2\sqrt{3}}{27y^{3/2}}\right)\ln\left(-\sqrt{3y}+y+1\right) - \frac{4}{3y} + \frac{4\beta}{27}\frac{\left(4y^2-4y+1\right)}{\left(y^2-y+1\right)y} + \\ &\left(\frac{9y^{3/2}+3y\sqrt{3}+3y^2\sqrt{3}-\sqrt{3}}{9y^{3/2}} + \beta\frac{-12y\sqrt{3}+6y^2\sqrt{3}-9y^{3/2}-2\sqrt{3}}{27y^{3/2}}\right)\ln\left(\sqrt{3y}+y+1\right) + \frac{13}{27}\beta\ln\left(y^2-y+1\right) \end{aligned} \tag{27}$$

Next, the macroscopic stresses at yielding are obtained by derivation (i.e. substitute Eq. (27) in Eq. (19)).

Thus, the yield surface of the porous aggregate for $\Sigma_m \geq 0$ and $J_3^{\Sigma} \leq 0$ is:

$$\begin{aligned} \left|\Sigma_{11}-\Sigma_{33}\right| &= \Sigma_e = -\frac{\sqrt{3}u^2}{4B}\left(\frac{1}{f}F_1'\left(\frac{u}{f}\right)-F_1'(u)\right) \\ \Sigma_m &= 2\Sigma_{11}+\Sigma_{33} = \frac{2}{3}\frac{\sqrt{3}u^2}{4B}\left(F_1\left(\frac{u}{f}\right)-F_1(u)+u\left(\frac{1}{f}F_1'\left(\frac{u}{f}\right)-F_1'(u)\right)\right) \end{aligned} \quad (28)$$

with $F_1'(y)$ being the derivative of the function $F_1(y)$ defined by Eq. (27).

(ii) For stress states such that $\Sigma_m \geq 0$ and $J_3^{\Sigma} > 0$ , substitution of the expression of the local plastic dissipation given by Eq. (23) into Eq. (18) and the change of variable given by Eq. (25), leads to the following expression of the overall plastic dissipation of the porous solid,

$$\Pi^+(\mathbf{D}, f) = \frac{\sqrt{3}}{4B}(uD_e)\int_u^{u/f}\int_{-1}^{1}\left(\sqrt{\left(y^2-y\left(3\alpha^2-1\right)+1\right)}+\frac{\beta}{27}\frac{(y-1)^2\left(2-y\left(9\alpha^2-5\right)+2y^2\right)^2}{\left(y^2+y\left(3\alpha^2-1\right)+1\right)^{5/2}}\right)\frac{dy}{y^2}d\alpha \quad (29)$$

Further integration of the above equation leads to the analytical expression of the overall plastic dissipation $\Pi^+(\mathbf{D}, f)$, which in this case has three branches:

$$\begin{cases} \Pi^+(\mathbf{D}, f) = \frac{\sqrt{3}uDe}{4B}\left(G_1\left(\frac{u}{f}\right)-G_1(u)\right), \text{ if } u<f \\ \Pi^+(\mathbf{D}, f) = \frac{\sqrt{3}uDe}{4B}\left(G_2\left(\frac{u}{f}\right)-G_1(u)+\beta\left(-\frac{7}{27}\pi-\frac{4\sqrt{3}}{27}\ln(3)+\frac{4\sqrt{3}}{9}\right)-\sqrt{3}\ln(3)-\frac{4\sqrt{3}}{3}+\frac{\pi}{9}\right), \text{ if } f<u<1 \\ \Pi^+(\mathbf{D}, f) = \frac{\sqrt{3}uDe}{4B}\left(G_2\left(\frac{u}{f}\right)-G_2(u)\right), \text{ if } u>1 \end{cases} \quad (30)$$

with

$$G_1(y) = -\frac{2\beta}{\sqrt{3}}\arctan\left(\frac{2y+1}{\sqrt{3}}\right) - \frac{2\sqrt{3}}{27}(3+11\beta)\left(\arctan\left(\frac{2\sqrt{y}-1}{\sqrt{3}}\right) - \arctan\left(\frac{2\sqrt{y}+1}{\sqrt{3}}\right)\right) +$$

$$-\frac{2\sqrt{3}}{9y^{3/2}}\left(3y^2-3y-1+\frac{2\beta}{3}\left(3y^2+6y-1\right)\right)\arctan\left(\frac{\sqrt{3y}}{y-1}\right) - \frac{(4\beta+27)}{27}\ln\left(y^2+y+1\right) + \tag{31}$$

$$\left(\frac{4}{27}\frac{\beta\left(4y^2+4y+1\right)-9\left(y^2+y+1\right)}{y\left(y^2+y+1\right)}\right),$$

and

$$G_2(y) = -\frac{2}{\sqrt{3}}\beta\arctan\left(\frac{2y+1}{\sqrt{3}}\right) + \frac{2\sqrt{3}}{27}(3+11\beta)\left(\arctan\left(\frac{2\sqrt{y}-1}{\sqrt{3}}\right) - \arctan\left(\frac{2\sqrt{y}+1}{\sqrt{3}}\right)\right)$$

$$+\frac{2\sqrt{3}}{9y^{3/2}}\left(3y^2-3y-1+\frac{2\beta}{3}\left(3y^2-1+6y\right)\right)\arctan\left(\frac{\sqrt{3y}}{y-1}\right) + \frac{27+4\beta}{27}\ln\left(y^2+y+1\right) \tag{32}$$

$$-\frac{4}{27}\frac{\beta\left(4y^2+4y+1\right)-9y-9-9y^2}{y\left(y^2+y+1\right)}$$

Hence, for stress states such that $\Sigma_m \geq 0$ and $J_3^\Sigma > 0$ the yield surface of the porous solid is:

$$\begin{cases} \Sigma_e = -\frac{\sqrt{3}u^2}{4B}\left(\frac{1}{f}G_1'\left(\frac{u}{f}\right) - G_1'(u)\right);\ \Sigma_m = \frac{2}{3}\frac{\sqrt{3}u^2}{4B}\left(G_1\left(\frac{u}{f}\right) - G_1(u) + u\left(\frac{1}{f}G_1'\left(\frac{u}{f}\right) - G_1'(u)\right)\right),\ \forall\ u<f \\ \Sigma_e = -\frac{\sqrt{3}u^2}{4B}\left(\frac{1}{f}G_1'\left(\frac{u}{f}\right) - G_2'(u)\right);\ \Sigma_m = \frac{2}{3}\frac{\sqrt{3}u^2}{4B}\left(\begin{array}{c} G_2\left(\frac{u}{f}\right) - G_1(u) + \beta\left(-\frac{7}{27}\pi - \frac{4\sqrt{3}}{27}\ln(3) + \frac{4\sqrt{3}}{9}\right) - \\ \sqrt{3}\ln(3) - \frac{4\sqrt{3}}{3} + \frac{\pi}{9} + u\left(\frac{1}{f}G_1'\left(\frac{u}{f}\right) - G_1'(u)\right) \end{array}\right),\ \forall\ f<u<1 \\ \Sigma_e = -\frac{\sqrt{3}u^2}{4B}\left(\frac{1}{f}G_2'\left(\frac{u}{f}\right) - G_2'(u)\right);\ \Sigma_m = \frac{2}{3}\frac{\sqrt{3}u^2}{4B}\left(G_2\left(\frac{u}{f}\right) - G_2(u) + u\left(\frac{1}{f}G_2'\left(\frac{u}{f}\right) - G_2'(u)\right)\right),\ \forall\ u>1 \end{cases} \tag{33}$$

with $G_1'(y)$ and $G_2'(y)$ denoting the derivatives of the function $G_1(y)$ and $G_2(y)$, respectively.

(iii) Based on the centro-symmetry property of the yield locus (Eq. (18)), demonstrated earlier, the parametric representation of the yield locus corresponding to stress states such that $\Sigma_m \leq 0$ , $J_3^\Sigma \geq 0$ can be easily obtained from Eq. (28) and Eq.(33), respectively.

Thus, for stress states such that $\Sigma_m \leq 0$ and $J_3^\Sigma \geq 0$:

$$\begin{cases} \dfrac{\Sigma_m}{\sigma_T} = -\left.\dfrac{\Sigma_m}{\sigma_T}\right| J_3^\Sigma \leq 0, \Sigma_m \geq 0 \\ \\ \dfrac{\Sigma_e}{\sigma_T} = \left.\dfrac{\Sigma_e}{\sigma_T}\right| J_3^\Sigma \leq 0, \Sigma_m \geq 0 \end{cases}, \tag{34}$$

the right-hand expressions of Eq. (34) being given by the corresponding Eq. (28);

(iv) For $\Sigma_m \leq 0$ and $J_3^\Sigma \leq 0$, the yield surface is:

$$\begin{cases} \dfrac{\Sigma_m}{\sigma_T} = -\left.\dfrac{\Sigma_m}{\sigma_T}\right| J_3^\Sigma \geq 0, \Sigma_m \geq 0 \\ \\ \dfrac{\Sigma_e}{\sigma_T} = \left.\dfrac{\Sigma_e}{\sigma_T}\right| J_3^\Sigma \geq 0, \Sigma_m \geq 0 \end{cases} \tag{35}$$

where the right-hand side expressions of Eq. (35) are given by the corresponding Eq. (33).

It is worth noting that the yield surface of the porous material is smooth. In particular, there are no singularities for hydrostatic states: $\lim\limits_{\substack{u\to\infty \\ J_3^\Sigma \leq 0}} \left(\Sigma_m\right) = \lim\limits_{\substack{u\to\infty \\ J_3^\Sigma \geq 0}} \left(\Sigma_m\right) = -\frac{2}{3}\sigma_0 \ln f$ . Furthermore, it is very interesting to note that the predicted yield limit for purely hydrostatic tensile loadings is the same as the yield limit for purely hydrostatic compression loadings, and it coincides with the yield limit under hydrostatic loadings of a porous solid with matrix obeying the von Mises, and Tresca criterion, respectively. Indeed, the predicted yield limit for hydrostatic tensile loadings is obtained by taking the limit when $u \to \infty$ of Eq. (28), and Eq.(33), respectively:

$$\lim_{\substack{u\to\infty \\ J_3^\Sigma\le 0}}\left(\Sigma_m\right)=\lim_{\substack{u\to\infty \\ J_3^\Sigma\ge 0}}\left(\Sigma_m\right)=-\frac{2}{3}\sigma_0\ln f \text{ and } \lim_{\substack{u\to\infty \\ J_3^\Sigma\le 0}}\left(\Sigma_e\right)=\lim_{\substack{u\to\infty \\ J_3^\Sigma\ge 0}}\left(\Sigma_e\right)=0 \qquad (36)$$

For purely deviatoric axisymmetric loadings, yielding occurs at $\Sigma_e=\sigma_0\left(1-f\right)$, irrespective of the sign of $J_3^\Sigma$. Indeed, by taking the limit $u\to 0$ in Eq. (28) and Eq.(33), respectively, we obtain:

$$\lim_{\substack{u\to 0 \\ J_3^\Sigma\le 0}}\left(\Sigma_m\right)=\lim_{\substack{u\to 0 \\ J_3^\Sigma\ge 0}}\left(\Sigma_m\right)=0 \text{ and } \lim_{\substack{u\to 0 \\ J_3^\Sigma\le 0}}\left(\Sigma_e\right)=\lim_{\substack{u\to 0 \\ J_3^\Sigma\ge 0}}\left(\Sigma_e\right)=\sigma_0\left(1-f\right), \qquad (37)$$

which is the yield limit for axisymmetric purely deviatoric loadings for both a porous Tresca and a porous von Mises material.

If the parameter $\beta$=0, the new criterion for porous materials derived in this paper (Eq. (28) and Eq. (33)) reduces to the criterion for porous solids with von Mises matrix developed by Cazacu et al. (2013). It is important to note that in the present development as well as in Cazacu et al. (2013) no approximations were made when calculating the local plastic dissipation of the respective porous material. Contrary to Gurson (1975; 1977), the cross-term $D_m D_e$ involved in the expression of the local plastic dissipation $\pi(\mathbf{d})$ (see Eq. (22)-(23)) was not neglected. Neglecting this coupling term would have resulted in erasing the specificities of the plastic deformation of the matrix and the sensitivity to $J_3^\Sigma$ of the resulting yield criterion of the porous solid.

## 4. Role played by the matrix sensitivity to the third-invariant on the mechanical response of the porous solid

### 4.1. Effect of the matrix sensitivity to the third-invariant on yielding of the porous aggregate

We begin by investigating the role played by the matrix sensitivity to both invariants of plastic deformation on yielding of the porous solid. Fig. 3 shows the representation in the plane ($\Sigma_m/\sigma_0$, $\left|\Sigma_{11}-\Sigma_{33}\right|/\sigma_0$) of the yield surface according to the criterion developed (see Eq. (28) and Eq. (33)) for a porous metallic material with matrix characterized by a value of the parameter $\beta$=0.38 and a porosity f = 5% (see Fig.3). Let denote by $T=\left(\Sigma_m/\Sigma_e\right)=\left(\Sigma_m/\left|\Sigma_{11}-\Sigma_{33}\right|\right)$, the stress triaxiality. As mentioned, for axisymmetric loadings, the equivalent stress $\Sigma_e$ reduces to $\left|\Sigma_{11}-\Sigma_{33}\right|$, the mean stress is: $\Sigma_m=\left(2\Sigma_{11}+\Sigma_{33}\right)/3$, and the third-invariant of the stress deviator is: $J_3^\Sigma=-\dfrac{2}{27}\left(\Sigma_{11}-\Sigma_{33}\right)^3$. Thus, for loadings such that the axial stress is smaller than the lateral stress, i.e. $\Sigma_{33}\leq\Sigma_{11}=\Sigma_{22}$, we have: $J_3^\Sigma\leq 0$, while for loadings such that the axial stress is larger than the lateral stress, i.e. $\Sigma_{11}=\Sigma_{22}\leq\Sigma_{33}$, we have $J_3^\Sigma\geq 0$. According to the new criterion for porous solids (see also Eq. (33) and Eq. (34)) there are only two types of loadings for which $J_3^\Sigma$ has no effect on yielding: purely deviatoric (T=0) and purely hydrostatic (T = $\pm\infty$). For all other loadings, yielding depends on $J_3^\Sigma$. For β=0.38, under tensile loading ($\Sigma_m\geq 0$) the response for loadings such that $J_3^\Sigma\geq 0$ is softer than that for loadings corresponding to $J_3^\Sigma\leq 0$ (the curve corresponding to $J_3^\Sigma\geq 0$ is below that corresponding to $J_3^\Sigma\leq 0$) while under compressive loadings ($\Sigma_m\leq 0$) the reverse occurs. Furthermore, the yield point corresponding to a given stress-triaxiality T and $J_3^\Sigma$ >0 is symmetric, with respect to the vertical axis, $\Sigma_m$ = 0, to the point corresponding to (-T) and ($J_3^\Sigma$ ≤0). This confirms that the yield locus is centro-symmetric, satisfying the invariance properties given by Eq. (15) Note that while for purely deviatoric loading, the response is the same irrespective of the sign of $J_3^\Sigma$, the effect of $J_3^\Sigma$ becomes stronger with increasing stress-triaxiality (see in Fig. 3 the different zooms of the yield surface in the tensile and compressive sectors, respectively).

In the rest of this paper, we will take advantage of the fact that irrespective of the value of the parameter β the yield surface of the porous solid is centro-symmetric, and represent and analyze only the quadrant of the yield surface defined by ($\Sigma_e$, $\Sigma_m$) with $\Sigma_m$ >0. In Fig. 4 are represented for the same level of porosity (f=0.05), the yield surfaces corresponding to materials with matrix characterized by β=0.38, 0.2, 0 (Von Mises matrix), -0.15 and -0.35, respectively. It is very interesting to note that if the matrix is characterized by $\beta\geq 0$ i.e. its plastic potential is exterior or coincides with von Mises (see also Fig.1), the response of the porous material for loadings at $J_3^\Sigma \geq 0$ is softer than that for loadings at $J_3^\Sigma \leq 0$ (yield curve corresponding to $J_3^\Sigma \geq 0$ is below that corresponding to $J_3^\Sigma \leq 0$). The stronger the sensitivity of the matrix plastic deformation to the third-invariant (i.e. larger the value of β), the stronger is the influence of $J_3^\Sigma$ on yielding of the porous solid (see Fig. 4(a)-4(c). It is very interesting to note that for the material with matrix characterized by $\beta = -0.15$ and porosity f =0.05 there is practically no influence of $J_3^\Sigma$ on the behavior (see Fig. 4(d) showing that the yield surface corresponding to $J_3^\Sigma \leq 0$ almost coincides with the yield surface that corresponds to $J_3^\Sigma \geq 0$). It means that although the matrix behavior depends on both invariants, the presence of voids practically erases the influence of $J_3^\Sigma$ on the yielding of the void-matrix aggregate. The same conclusion, i.e. practically no influence of $J_3^\Sigma$ on the response of the porous solid, applies to a material characterized by a porosity f = 0.01 and matrix with $\beta = -0.15$ (see Fig. 5 (c)). It is to be noted that the particular value of β, say $\beta_*$, that characterizes the matrix of the porous material that displays no influence on the third-invariant can be determined numerically by making use of Eq. (28) and Eq. (33), respectively. For fixed values of the porosity f, ranging from $10^{-5}$ to 0.15, $\beta_*$ is between –0.179 and –0.172.

### 4.2. Effect of the matrix sensitivity to the third-invariant on porosity evolution

Next, the role of the plastic flow of the matrix on void evolution is investigated. Given that the new potential for porous solids given by Eq. (28) and Eq. (33) accounts for the combined effects of the mean stress $\Sigma_m$ and the third-invariant $J_3^\Sigma$, void evolution should be affected by $J_3^\Sigma$. However, the value of the parameter β which describes the sensitivity of the plastic deformation of the matrix to the third-invariant ought to affect the rate of void evolution. This is clearly seen by comparing the predictions of the void growth versus the equivalent plastic strain Ee corresponding to materials with matrix characterized by β=0.38, 0.2, -0.15 and -0.35 (see Fig. 6) when subjected to axisymmetric loadings at fixed triaxiality T=1.5 and either $J_3^\Sigma < 0$ ($\Sigma_1 = \Sigma_2 \geq \Sigma_3$) or $J_3^\Sigma \geq 0$ ($\Sigma_1 = \Sigma_2 \leq \Sigma_3$). For all porous materials considered the initial porosity is: $f_0 = 0.005$. As discussed in the previous section, for the material characterized by a porosity $f_0$=0.005 and matrix with β= –0.15, close to $\beta_*$, there is practically no effect of $J_3^\Sigma$ on the yield surface (see

Fig. 4(d)). Consequently, the rates of void growth for $J_3^\Sigma \geq 0$ and $J_3^\Sigma < 0$ are almost the same (see Fig. 6 (c)). It is worth noting that if the matrix is characterized by β>β*, the rate of void growth is faster for $J_3^\Sigma \geq 0$ than for $J_3^\Sigma < 0$. The larger is the value of β, the stronger is the effect of $J_3^\Sigma$ on void growth (compare Fig. 6(a) which corresponds to a material with matrix characterized β =0.2 with Fig. 6(b) which corresponds to a material with β=0.38).

On the contrary, if the matrix is characterized by β <β*, the rate of void growth is faster for $J_3^\Sigma \leq 0$ than for $J_3^\Sigma \geq 0$ (see for example, Fig. 6(d) which presents the void evolution for the material with matrix characterized by β=–0.35). Because the new criterion is centro-symmetric ( $F\left(\Sigma_m, \Sigma_e, J_3^\Sigma, f\right) = F\left(-\Sigma_m, \Sigma_e, -J_3^\Sigma, f\right)$), the following conclusions can be drawn concerning the effect of the matrix sensitivity to the third-invariant of the stress-deviator $J_3^\Sigma$ on void collapse (i.e. void evolution for compressive mean stress):

- If the porous solid has the matrix characterized by β > β*, the rate of void collapse is faster for loadings at $J_3^\Sigma < 0$ than for loadings at $J_3^\Sigma \geq 0$;
- If the porous solid has the matrix characterized by β close to β*, there is practically no effect of $J_3^\Sigma$ on void closure;
- If the porous solid has the matrix characterized by a value of the parameter β < β* the rate of void closure is faster for loadings at $J_3^\Sigma \geq 0$ than for loadings at $J_3^\Sigma \leq 0$.

**5. Effect of the plastic flow of the matrix on the dilatational response: comparison between the mechanical response according to the new model and that of a porous Mises and porous Tresca material**

For most fully-dense isotropic metallic materials, the potential that governs their plastic behavior depends on both invariants of the strain-rate tensor. One of the objectives of this paper is to study how the deviations in the matrix plastic deformation from von Mises criterion or Tresca criterion influence the rate at which porosity accumulates. To this end, we compare the yield surface and

void evolution according to the new criterion for porous materials (Eq. (28) and Eq. (33)) with the predictions of Cazacu et al (2013) criterion for a material with matrix described by Von Mises criterion (β=0), and the predictions of Cazacu et al. (2014) criterion for a porous Tresca material.

We begin by comparing the cross-sections of the yield surfaces for porous solids containing 5% voids and with matrix characterized by Tresca, von Mises, and the new criterion corresponding to β=0.38, 0.2, -0.15 and -0.35. For loadings corresponding to $J_3^{\Sigma} \leq 0$ ($\Sigma_{11} = \Sigma_{22} \geq \Sigma_{33}$), the respective surfaces are shown in Fig. 7(a),8(a),9(a), 10(a) ; for loadings corresponding to $J_3^{\Sigma} \geq 0$ ($\Sigma_{11} = \Sigma_{22} \leq \Sigma_{33}$) the respective surfaces are shown in Fig. 7(b),8(b),9(b),10(b). It is very interesting to note that if the matrix is characterized by β >0, the yield surface of the porous solid lies between the yield surface of a porous Mises material and the yield surface of the porous Tresca solid. Specifically, for β >0 the porous Tresca yield surface is a lower bound while the porous von Mises surface is an upper bound. This is to be expected since for β >0, the plastic potential of the matrix lies between that of von Mises and Tresca potentials (see Fig.1). Moreover, the stronger the sensitivity to the third-invariant of the matrix (i.e. the larger the value of the parameter β ), the closer the yield surface is to that of a porous Tresca material (see Fig. 7-8).

On the other hand, if the matrix is characterized by β <0, the response of the porous solid is harder than that of a porous solid with von Mises matrix (β =0). The smaller the value of β is, the more pronounced is the difference in response as compared to that of a porous von Mises material. For example, compare Fig. 9-10 which show comparisons between the yield loci for the porous Mises material and those corresponding to β = -0.15 and β=-0.35, respectively. Note that the yield limit for purely deviatoric states and purely hydrostatic states is the same for all porous materials irrespective of the criterion governing the plastic deformation of the matrix.

In the following, the effect of the particularities of the plastic flow of the matrix on void evolution will be investigated.

As discussed in Section 4.2. when the value of the parameter β is close to $\beta_*$, the effect of the sign of $J_3^\Sigma$ on the void growth or closure rates is minimal. As an example, in Fig. 11-12 is shown the void evolution in a porous material with matrix characterized by $\beta = -0.15$ ( close to $\beta_* = -0.175$). At an equivalent plastic strain $E_e$=0.3, the void volume fraction for loading corresponding to $J_3^\Sigma > 0$ is almost identical with the void volume fraction corresponding to loadings at the same triaxiality and $J_3^\Sigma < 0$. (see Fig. 11 for predictions of void growth , and Fig. 12 for predictions of void collapse). Nevertheless, since for $\beta=-0.15<0$ the potential of the matrix is interior to that of von Mises (Fig.1) the void growth and void closure rates in the material are slower than in both a porous material with a von Mises matrix and a porous solid with Tresca matrix.

Next, we compare the predictions of the new criterion for porous solids with matrix characterized by β=0.38, 0.2, -0.15 and -0.35 with the void evolution in a material with von Mises matrix (β=0), and a Tresca matrix, respectively. Specifically, Fig. 13-14 compare the void volume fraction $f/f_0$ versus the effective plastic strain curves for materials with matrix characterized by β=0.38 and β=0.20, respectively subject to axisymmetric loading histories corresponding to either ( $J_3^\Sigma \le 0$ ) and ( $J_3^\Sigma \ge 0$ ) and fixed positive stress triaxiality T = 1.5. For all porous materials, the initial void volume fraction is $f_0$=0.005. Note that the combination between invariants in the matrix is such that β>0, so irrespective of the type of loading (i.e. sign of $J_3^\Sigma$), the rate of void growth is faster than in a porous von Mises material (β =0) and lower than in a porous Tresca material. As an example, note that for the material with matrix characterized by β=0.38 for loadings such that $J_3^\Sigma \le 0$, at an equivalent plastic strain of $E_e$=0.3 the void volume fraction is f= 7.82 $f_0$ in the porous Mises solid, 9.41 $f_0$ in the material with matrix characterized by β=0.38, against 11.29 $f_0$ in the porous Tresca solid. Comparison between Fig. 13(a) and Fig. 13(b) show that all the criteria predict an influence of $J_3^\Sigma$ on void growth, the rate of void growth being faster for $J_3^\Sigma \ge 0$ (see Fig. 13(b)) than for $J_3^\Sigma \le 0$ (see Fig. 13(a)). However, the influence of $J_3^\Sigma$ on the rate of void growth is more pronounced for the material with matrix characterized by $\beta$=0.38 (13% difference between $J_3^{\Sigma} \leq 0$ and $J_3^{\Sigma} \geq 0$ at $E_e$=0.3) than for the porous Tresca solid (8% difference at $E_e$=0.3) and the porous Mises solid (5% difference at $E_e$=0.3).

The same conclusions can be drawn from the analysis of the void evolution in the porous solid with matrix characterized by $\beta$=0.2 (see Fig. 14). However, the influence of $J_3^{\Sigma}$ on void growth is less pronounced. As an example, at $E_e$=0.3, the difference between loadings corresponding to $J_3^{\Sigma} \leq 0$ and $J_3^{\Sigma} \geq 0$ is of 13% for the material with matrix characterized by $\beta$=0.38 against 9.5% for the one corresponding to a matrix with $\beta = 0.2$ and only 5% for a material with matrix characterized by $\beta$=0 ( von Mises matrix). In general, for $\beta > \beta_*$ the influence of $J_3^{\Sigma}$ on void growth is less pronounced as the value of the parameter $\beta$ decreases.

Since both the new criterion developed in Section 4 and Cazacu et al (2014) criterion for a porous Tresca material display centro-symmetry, the same effects of the weighting of the invariants in the matrix (i.e. of $\beta$) on the rate of void closure should occur. Comparison between the predictions of void closure in the same materials with matrix characterized by $\beta > 0$ for axisymmetric loadings but fixed compressive triaxiality T=-1.5 corresponding to either $J_3^{\Sigma} \leq 0$ or $J_3^{\Sigma} \geq 0$ are shown in Fig. 15-16. The initial porosity was considered higher ($f_0$ = 0.05) such as to allow a larger range of plastic strain to develop prior to pore closure. Irrespective of the sign of the third invariant, the rate of void closure in the material with matrix characterized by $\beta$=0.38 is much faster than in the porous Mises solid, and only slightly slower than the rate of void closure in the porous solid with Tresca matrix (see Fig. 15(a) for $J_3^{\Sigma} \leq 0$ and Fig. 15(b) for $J_3^{\Sigma} \geq 0$). As in the case of a porous Mises or porous Tresca material, if $\beta > \beta_*$, the rate of void closure is faster for loadings at $J_3^{\Sigma} \leq 0$ than for loadings at $J_3^{\Sigma} \geq 0$ (e.g. compare Fig. 15(a) and Fig. 15(b)). However, the influence of $J_3^{\Sigma}$ on the rate of void evolution is more pronounced for the material with matrix characterized by $\beta$=0.38 (18% difference between $J_3^{\Sigma} \leq 0$ and $J_3^{\Sigma} \geq 0$ at $E_e$=0.3) than for the porous Tresca solid (13 % difference at $E_e$=0.3) and the porous Mises solid (8 % difference at $E_e$=0.3).

The same conclusions can be drawn by analyzing the rate of void closure in the material with matrix characterized by β=0.2 (see Fig. 16) as compared to that in the porous von Mises and porous Tresca material, respectively. However, the influence of $J_3^\Sigma$ on the rate of void closure is less pronounced than in the case when the matrix is characterized by β=0.38. As an example, at $E_e$=0.3, the difference between the porosity corresponding to loadings at $J_3^\Sigma \leq 0$ and $J_3^\Sigma \geq 0$ is of 18% for the material with matrix characterized by β=0.38, against 12% for the material characterized by β=0.2 and 8% for the material with β=0 (porous Mises solid). In general, for $\beta > \beta_*$, the influence of $J_3^\Sigma$ on void evolution decreases as the value of the parameter β decreases. As seen previously, for stress triaxialities T different from zero or infinity the response of the porous Tresca material is softer than that of the porous von Mises material which in turn is softer than that of a porous material with matrix characterized by β<0(see Fig. 10). As a consequence, for β<0, the void growth rate and void closure rate in such materials will be slower than the rate of void evolution in a porous Mises material (β=0) and a porous Tresca material, respectively. Indeed, let examine Fig. 17 which shows the void growth evolution in a material with matrix characterized by $\beta = -0.35$ for axisymmetric loadings at a fixed stress triaxiality T=1.5 corresponding to $J_3^\Sigma \leq 0$ and $J_3^\Sigma \geq 0$, respectively. Note that at an equivalent plastic strain Ee= 0.3, for axisymmetric loading corresponding to $J_3^\Sigma \leq 0$ in the material with β=-0.35 the void volume fraction is f= 6.30 $f_0$, against f= 11.3 $f_0$ in the porous Tresca solid and f= 7.82 $f_0$ in the porous Mises solid (β=0). It is also worth noting that while for the porous Mises and porous Tresca solid, the void growth rate is faster for loadings at $J_3^\Sigma \geq 0$ than for loadings at $J_3^\Sigma \leq 0$, for the material with matrix characterized by β=-0.35, the void growth rate is slower for $J_3^\Sigma \geq 0$ than for $J_3^\Sigma \leq 0$. Therefore, for this material with matrix characterized by β=-0.35 < $\beta_*$, the particular weighting between the invariants of the matrix plastic deformation has a very important influence on all aspects of the mechanical response of the porous solid. Not only it slows down the void growth rate as compared to a von Mises porous solid, but the influence of the sign of $J_3^\Sigma$ on void growth rate is reverted. Due to the centro-symetry of all criteria, it follows that the void closure rate will also be slower than in a porous von Mises material. The same conclusions, i.e.

that the void closure rate will be faster for $J_3^{\Sigma} > 0$ than for $J_3^{\Sigma} < 0$ can be drawn for any porous material with matrix characterized by $\beta < \beta_*$ (see Fig. 18).

## 6. Conclusions

The fundamental question that we posed and addressed in this paper concerns the role played by the relative weighting of the invariants of the plastic deformation of the matrix on the overall mechanical response of a porous isotropic metallic material. To this end, we first proposed a new isotropic potential for description of the plastic behavior of the matrix that depends on both invariants of the strain-rate deviator. The relative weight of the two invariants is described by a material parameter $\beta$. Depending on the sign of the parameter $\beta$, the new plastic potential for the matrix is either interior to the von Mises strain-rate potential ($\beta$ <0), coincides with it ($\beta$ = 0) or it is exterior to it. An analytical yield criterion for porous metallic materials with matrix described by this new potential was developed using a rigorous micromechanical analysis based on Hill-Mandel lemma. This ensures that the criterion developed is an upper-bound estimate of the exact plastic potential. The limit-analysis was conducted for both tensile and compressive axisymmetric loading scenarios and spherical void geometry. Contrary to the case when the matrix is governed by von Mises criterion, for an incompressible matrix obeying the new potential, the calculation of the plastic dissipation is very challenging. This is because the plastic response of the matrix depends on a combination of the invariants of the local rate of deformation, **d**. However, it was shown that all the integrals representing the overall plastic dissipation can be calculated analytically. No simplifying approximations were considered when estimating the local and overall plastic dissipation, respectively. A closed-form representation of the yield surface of the porous aggregate is obtained (see Eqs. (28); Eq.(33)). For $\beta$=0, i.e. von Mises matrix, the new criterion reduces to the porous Mises criterion of Cazacu et al. (2013).

It is worth summarizing the salient features of the new criterion for porous solids developed:

- Irrespective of the value of the parameter β, which describes the relative weighting of the invariants of the plastic deformation in the matrix, the dilatational response of the porous solid depends on the sign of mean stress, $\Sigma_m$ (tension-compression asymmetry). The yield locus is centro-symmetric.

- There exists a particular combination of invariants i.e. value of the parameter β, called $\beta_*$, for which there is practically no effect of $J_3^\Sigma$ on the response of the porous solid. In other words, although the plastic deformation of the matrix depends on both invariants, the presence of voids in the material erases the influence of $J_3^\Sigma$ on the yielding of the void-matrix aggregate. This value $\beta_*$ is negative.
- If the matrix is characterized by $\beta > \beta_*$, the response of the porous material for tensile loadings ($\Sigma_m \geq 0$) corresponding to $J_3^\Sigma \geq 0$ is softer than that for tensile axisymmetric loadings at $J_3^\Sigma \leq 0$. The stronger the sensitivity of the matrix plastic deformation to the third-invariant (i.e. the larger β is), the stronger is the influence of $J_3^\Sigma$ on the yielding of the porous solid.

- On the contrary, if the matrix is characterized by a value of the parameter $\beta<\beta_*$, for tensile loadings ($\Sigma_m \geq 0$), the response of the porous material is softer for loadings at $J_3^\Sigma \leq 0$ than at $J_3^\Sigma >0$.

As concerns void evolution,

- If the porous solid has the matrix characterized by $\beta > \beta_*$, the rate of void growth or collapse is faster for loadings at $J_3^\Sigma \leq 0$ than for loadings at $J_3^\Sigma \geq 0$;

- If the porous solid has the matrix characterized by a value of the parameter $\beta < \beta_*$ the rate of void growth or void closure is faster for loadings at $J_3^\Sigma \geq 0$ than for loadings at $J_3^\Sigma \leq 0$.
- If the porous solid has the matrix characterized by β close to $\beta_*$, there is practically no effect of $J_3^\Sigma$ on void evolution.

A noteworthy result is that the softest response corresponds to a porous material with matrix governed by Tresca's criterion. However, depending on the value of β the rate of void growth or collapse can be either faster or slower than that of a porous Mises material which corresponds to a matrix characterized by $\beta = 0$. Specifically, if the matrix is characterized by $\beta > 0$, the void growth rate is faster than in a porous Mises material; the larger the value of β , the fastest the rate of void evolution, which approaches the one in a porous Tresca material. If the matrix is characterized by $\beta < 0$, the void growth rate and void growth closure rate predicted by the new porous model are slower than in a porous Mises material ($\beta = 0$). The smaller the value of β, the slower the rate of void growth as compared to a von Mises porous material.

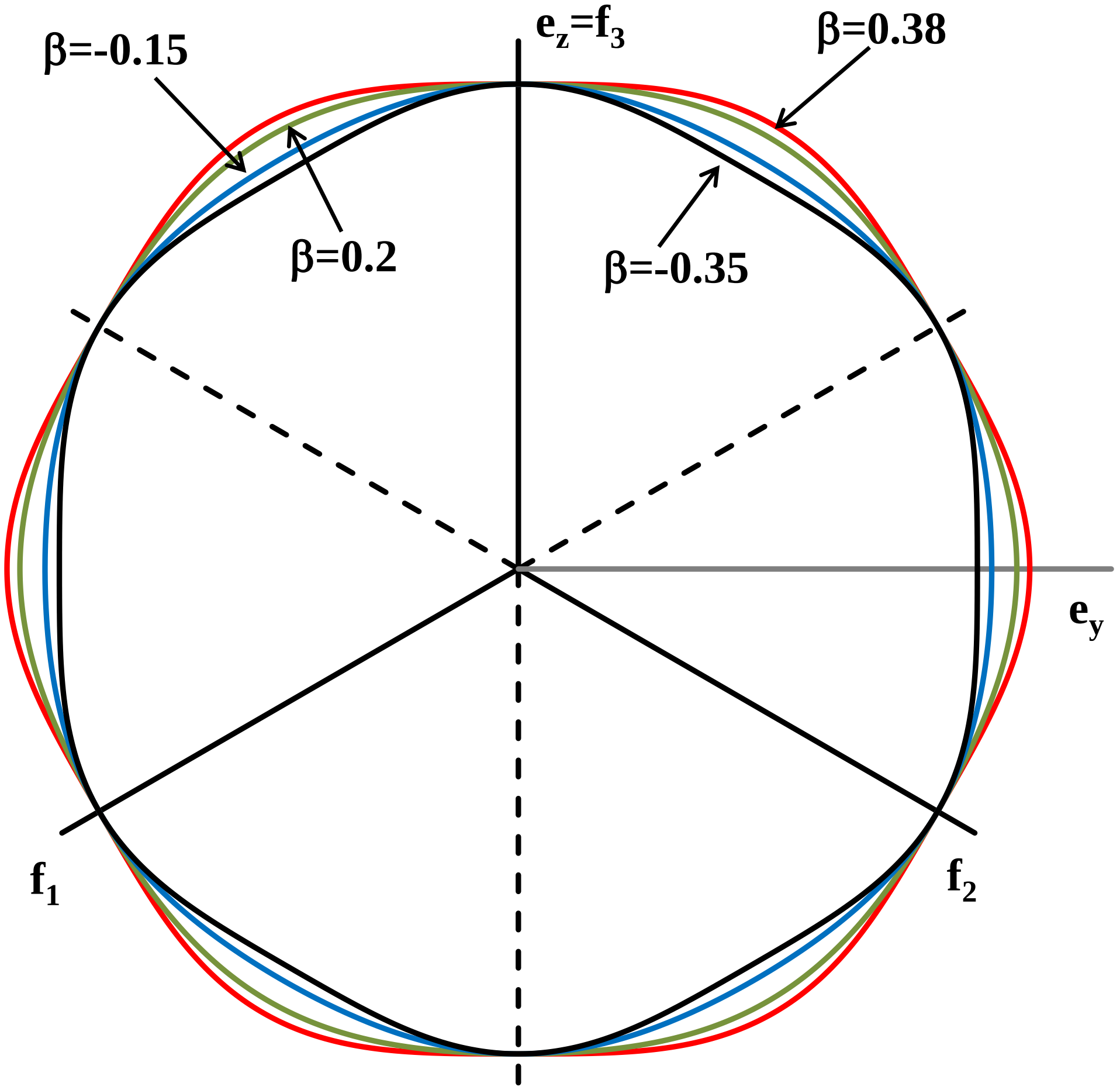

**Fig 1.** New strain rate potential for pressure-insensitive isotropic materials (Eq.(6)) in the octahedral plane for several values of the parameter β=0.38, 0.2, -0.15 and -0.35.

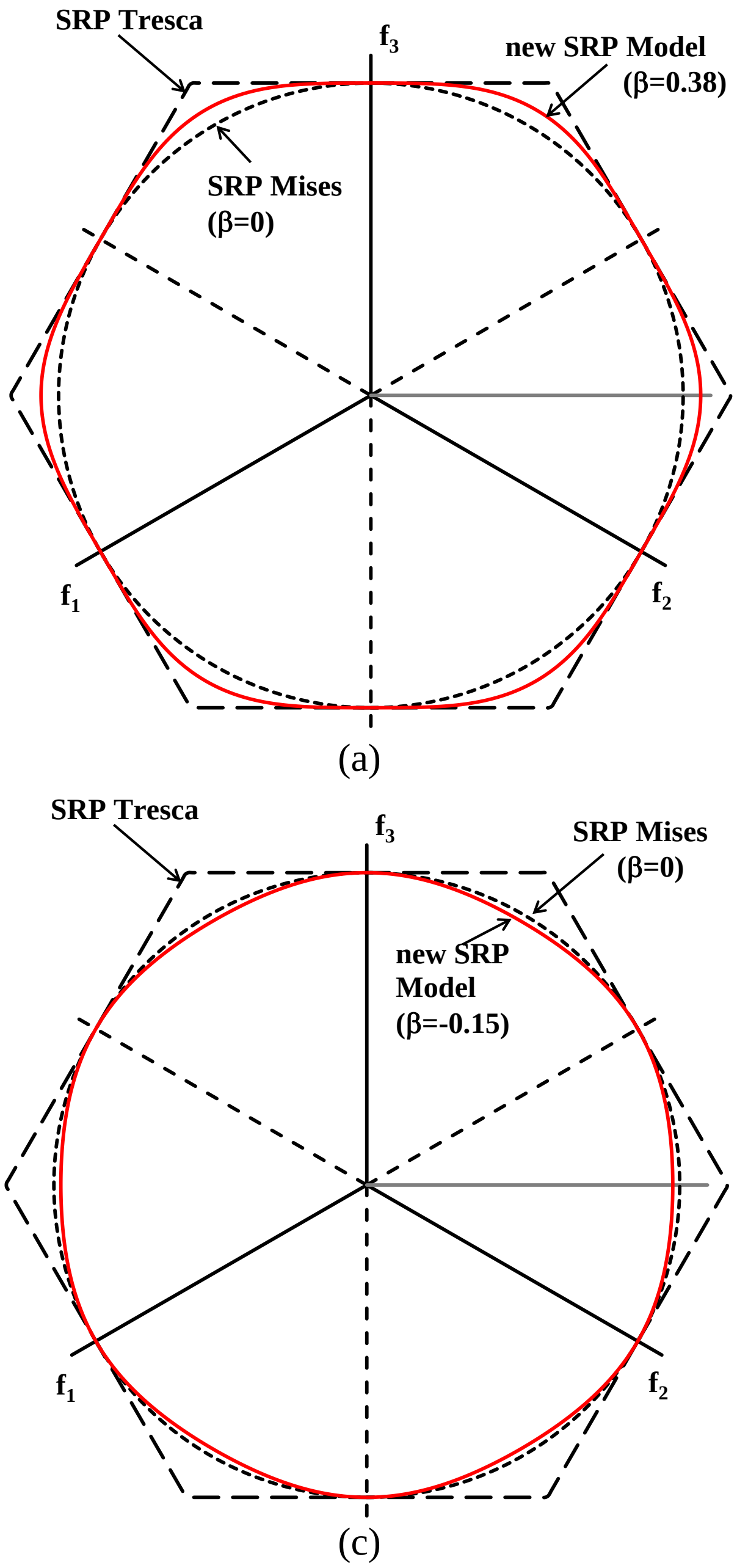

**Fig 2**. Representation in the octahedral plane and comparison of the von Mises strain-rate potential (dotted line), Tresca's strain-rate potential (dashed line), and the new strain-rate potential for pressure-insensitive isotropic materials (Eq.(6)) corresponding to several values of the parameter β: (a) β =0.38 (b) β=0.2, (c) β=–0.15; (d) β=-0.35.

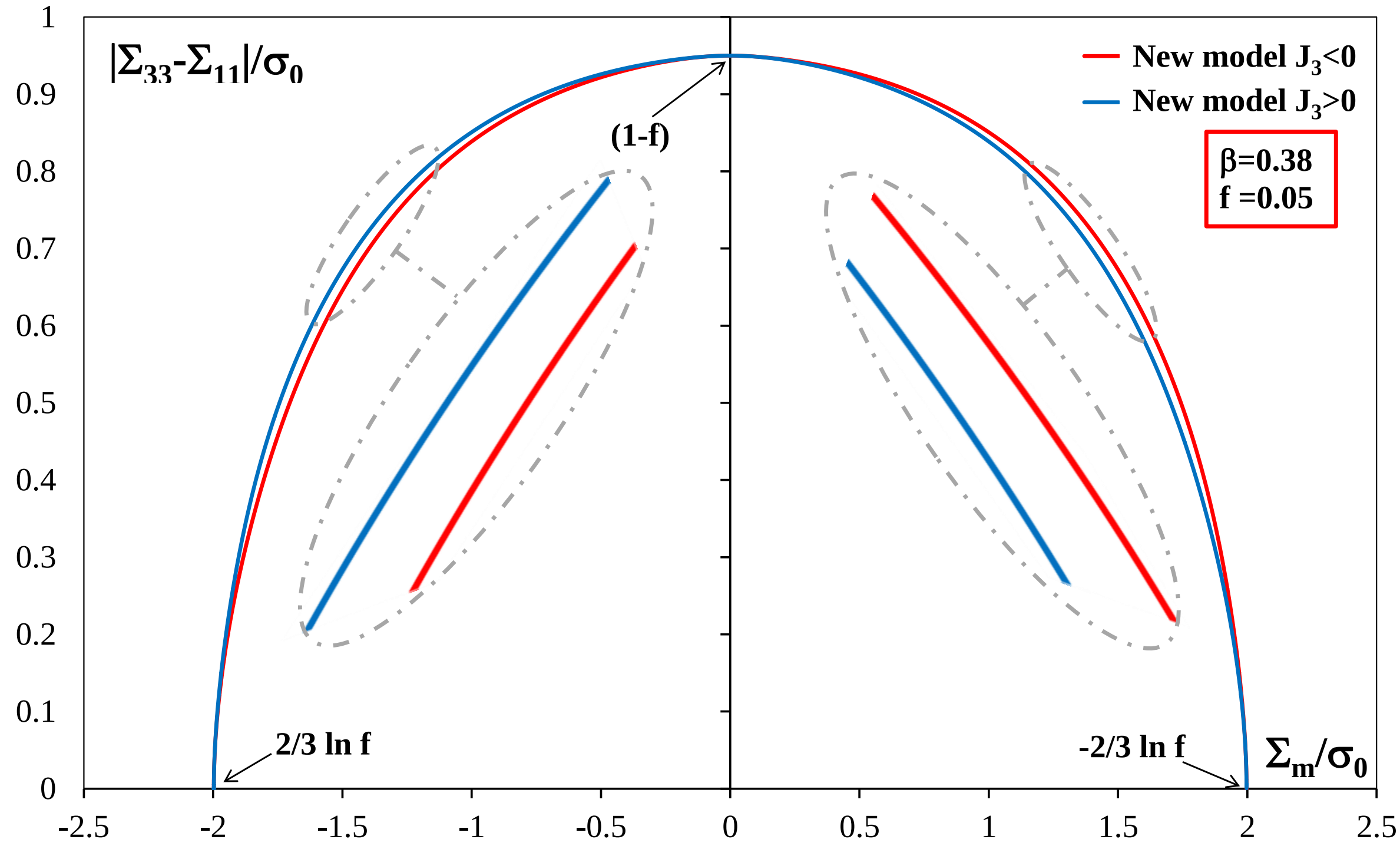

**Fig**. **3**. Yield surface according to the developed criterion for porous solids (Eq. (28)-(33)) for matrix characterized by a value of the parameter β=0.38 and random void distribution corresponding to void volume fraction f =0.05 for axisymmetric loadings corresponding to $J_3^{\Sigma} \leq 0$ ($\Sigma_{11} = \Sigma_{22} \geq \Sigma_{33}$) and $J_3^{\Sigma} \geq 0$ ($\Sigma_{11} = \Sigma_{22} \leq \Sigma_{33}$), respectively.

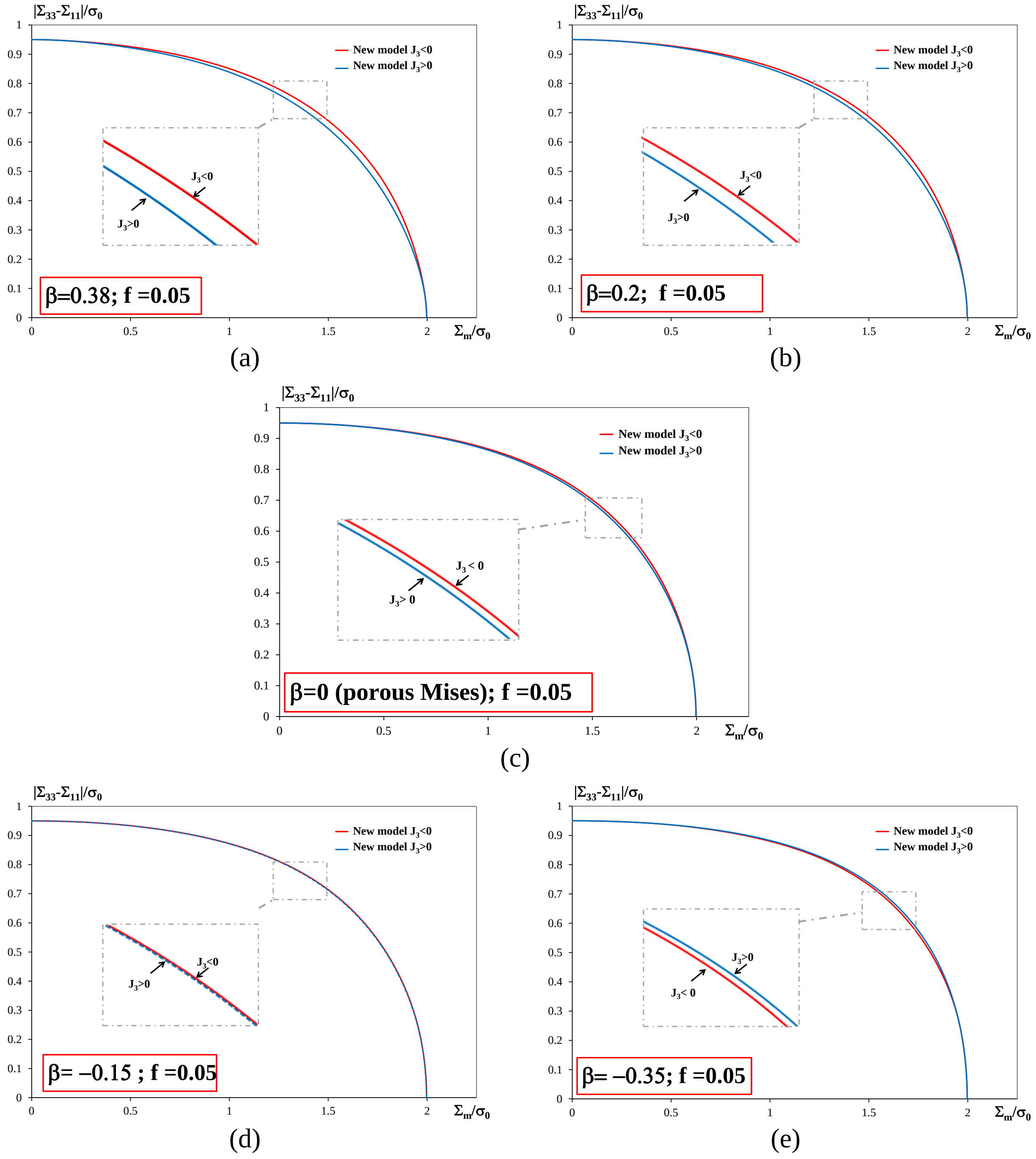

**Fig**. **4.** Yield surface according to the new criterion for porous solids (Eq. (28)-(33)) and matrix characterized by: (a) β=0.38; (b) β= 0.2 ; (c) β=0 (von Mises matrix ); (d) β=-0.15; (e) β=–0.35. For all materials the void volume fraction is f= 0.05.

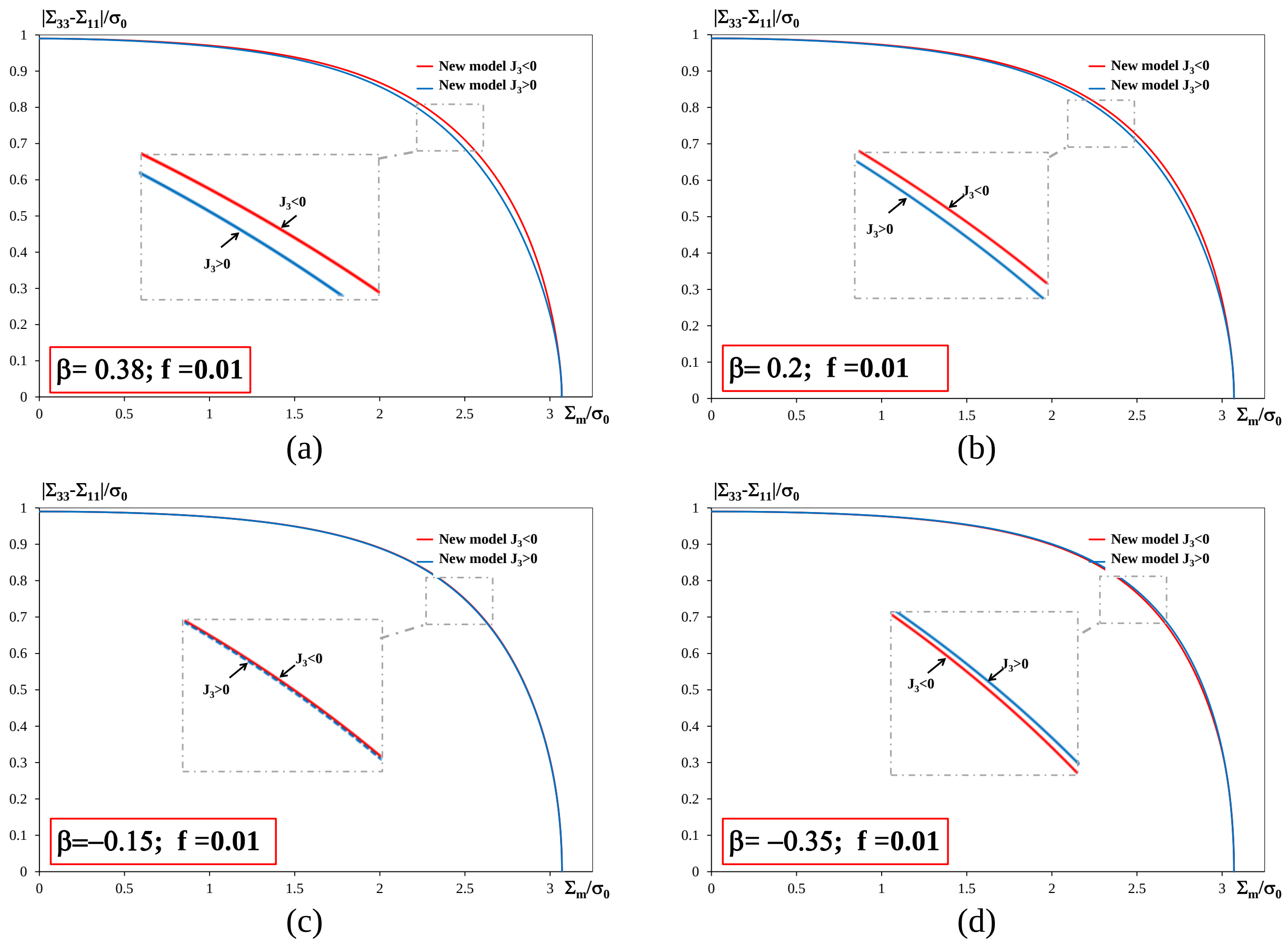

**Fig. 5.** Yield surface according to the new criterion for porous materials (Eq. (28)-(33)) with matrix characterized by: (a) β=0.38; (b) β= 0.2 ; (c) β=0 (von Mises matrix ); (d) β=-0.15; (e) β=–0.35. For all materials the void volume fraction is f=0.01.

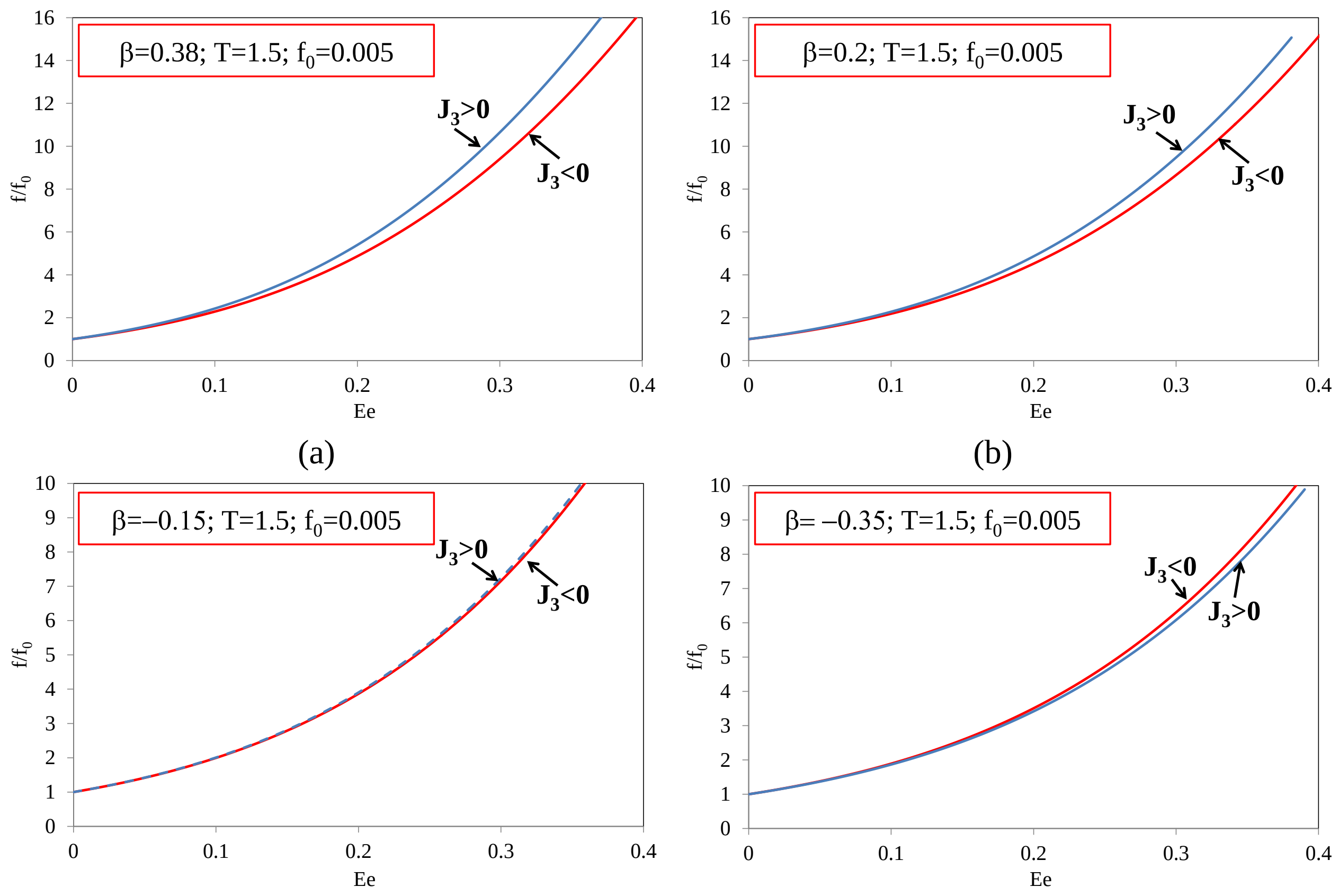

**Fig. 6.** Effect of the third-invariant $J_3^{\Sigma}$ on the void volume fraction $(f/f_0)$ evolution with the equivalent plastic strain for axisymmetric loadings at fixed triaxiality T=1.5 predicted by new criterion (Eq. (28)-(33)) for porous materials characterized by matrix with (a) β=0.38; (b) β= 0.2; (c) β= -0.15; (d) β= -0.35. For all materials the initial porosity is: $f_0 = 0.005$.

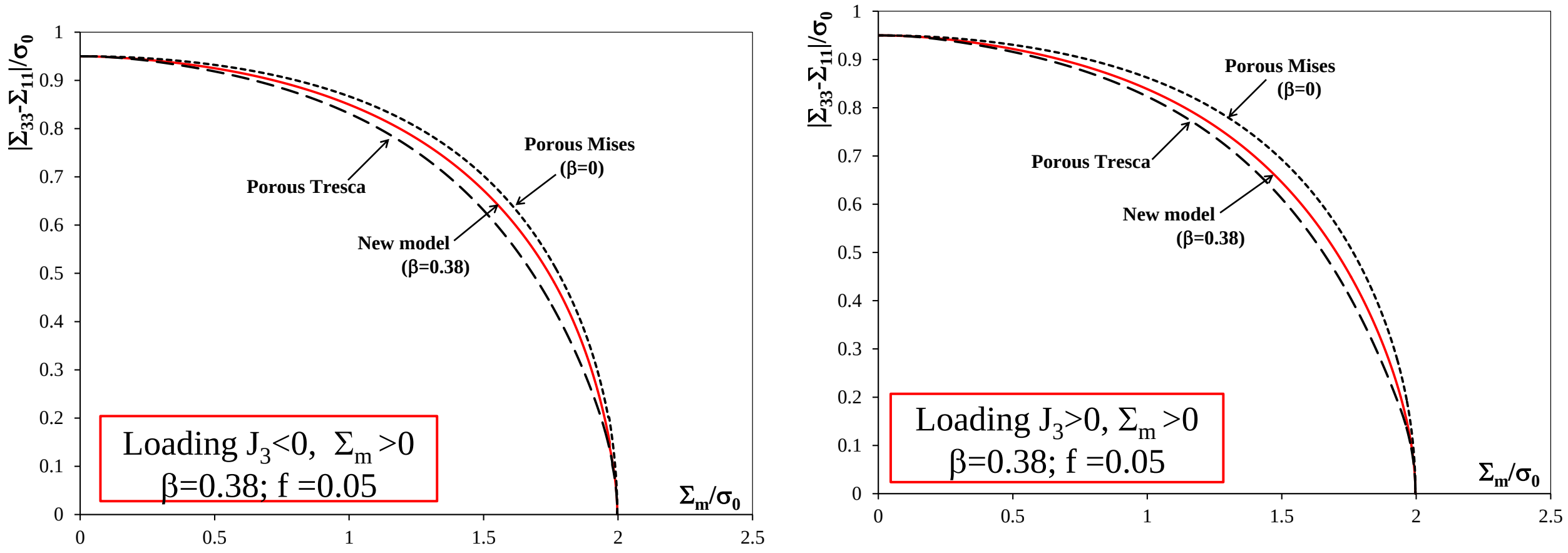

(a) (b)

**Fig. 7**. Comparison between the yield surfaces according to the criterion for porous solids with von Mises matrix of Cazacu et al. (2013), the porous Tresca criterion (Cazacu et al. 2014) and the new criterion for porous solid (Eq. (28)-(33)) for matrix with β=0.38, respectively for axisymmetric loadings such that: (a) $J_3^\Sigma \geq 0$ $(\Sigma_{11} = \Sigma_{22} \leq \Sigma_{33})$; (b) $J_3^\Sigma \leq 0$ $(\Sigma_{11} = \Sigma_{22} \geq \Sigma_{33})$ corresponding to the same porosity f=0.05.

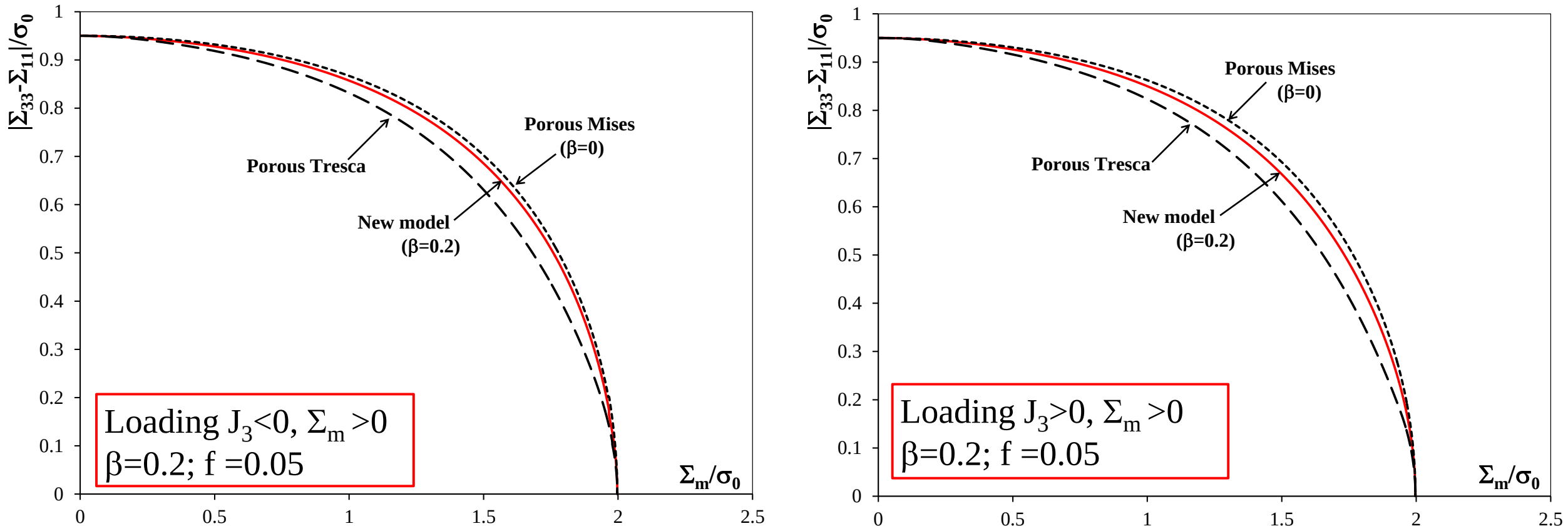

(a) (b)

**Fig. 8**. Comparison between the yield surfaces according to the criterion for porous solids with von Mises matrix of Cazacu et al. (2013), the porous Tresca criterion (Cazacu et al. 2014) and the new criterion for porous solid (Eq. (28)-(33)) for matrix with β=0.2, respectively for axisymmetric loadings such that: (a) $J_3^\Sigma \geq 0$ $(\Sigma_{11} = \Sigma_{22} \leq \Sigma_{33})$; (b) $J_3^\Sigma \leq 0$ $(\Sigma_{11} = \Sigma_{22} \geq \Sigma_{33})$ corresponding to the same porosity f=0.05.

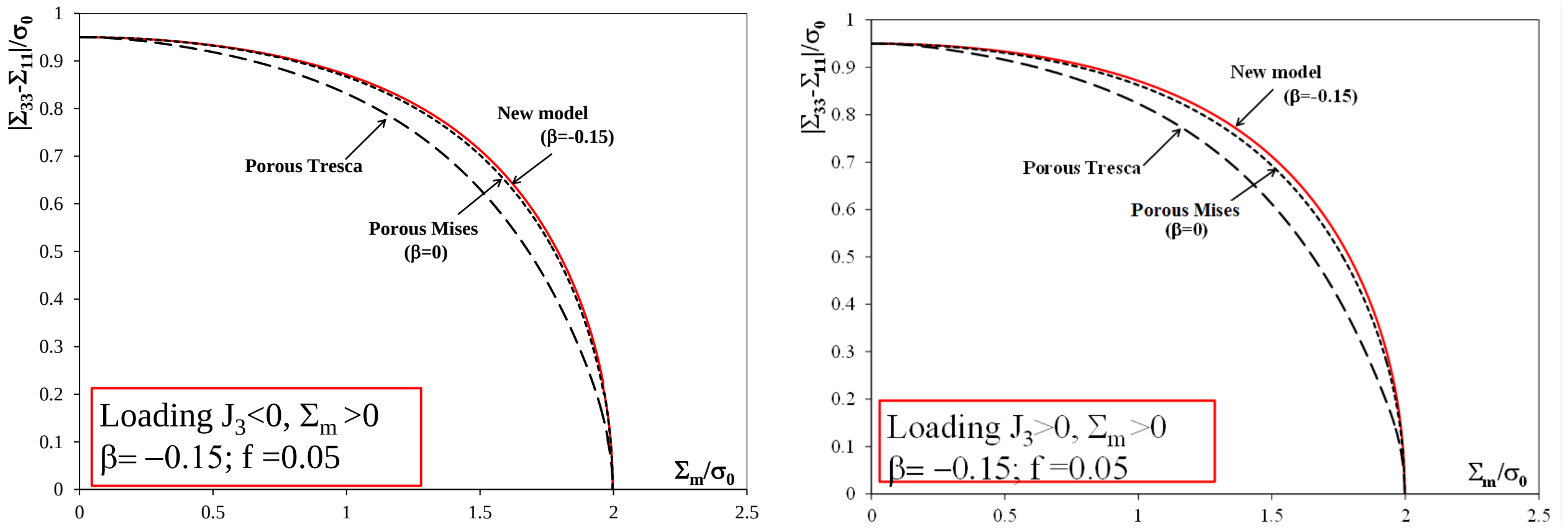

(a) (b)

**Fig. 9** Comparison between the yield surfaces according to the criterion for porous solids with von Mises matrix of Cazacu et al. (2013), the porous Tresca criterion (Cazacu et al. 2014) and the new criterion for porous solid (Eq. (28)-(33)) for matrix with β=-0.15, respectively for axisymmetric loadings such that: (a) $J_3^{\Sigma} \geq 0$ $(\Sigma_{11} = \Sigma_{22} \leq \Sigma_{33})$; (b) $J_3^{\Sigma} \leq 0$ $(\Sigma_{11} = \Sigma_{22} \geq \Sigma_{33})$ corresponding to the same porosity f=0.05.

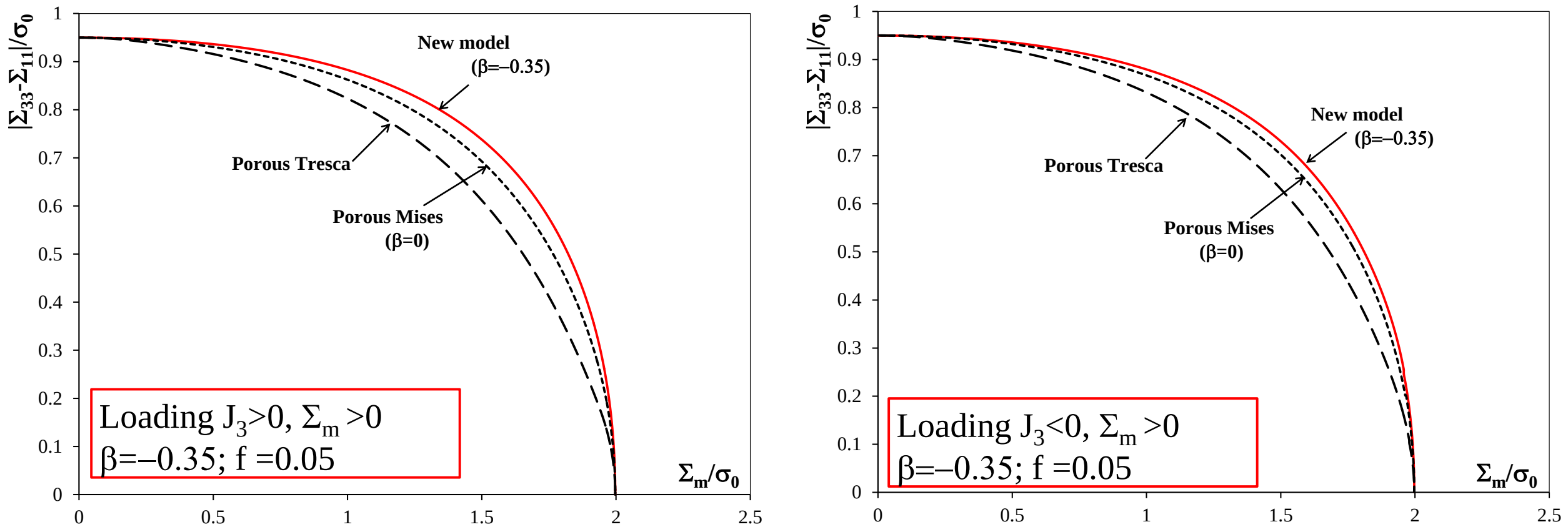

(a) (b)

**Fig. 10.** Comparison between the yield surfaces according to the criterion for porous solids with von Mises matrix of Cazacu et al. (2013), the porous Tresca criterion (Cazacu et al. 2014) and the new criterion for porous solid (Eq. (28)-(33)) for a material with matrix with β = -0.35, respectively for axisymmetric loadings such that: (a) $J_3^{\Sigma} \geq 0$ $(\Sigma_{11} = \Sigma_{22} \leq \Sigma_{33})$; (b) $J_3^{\Sigma} \leq 0$

$(\Sigma_{11} = \Sigma_{22} \geq \Sigma_{33})$ corresponding to the same porosity $f_0$=0.05.

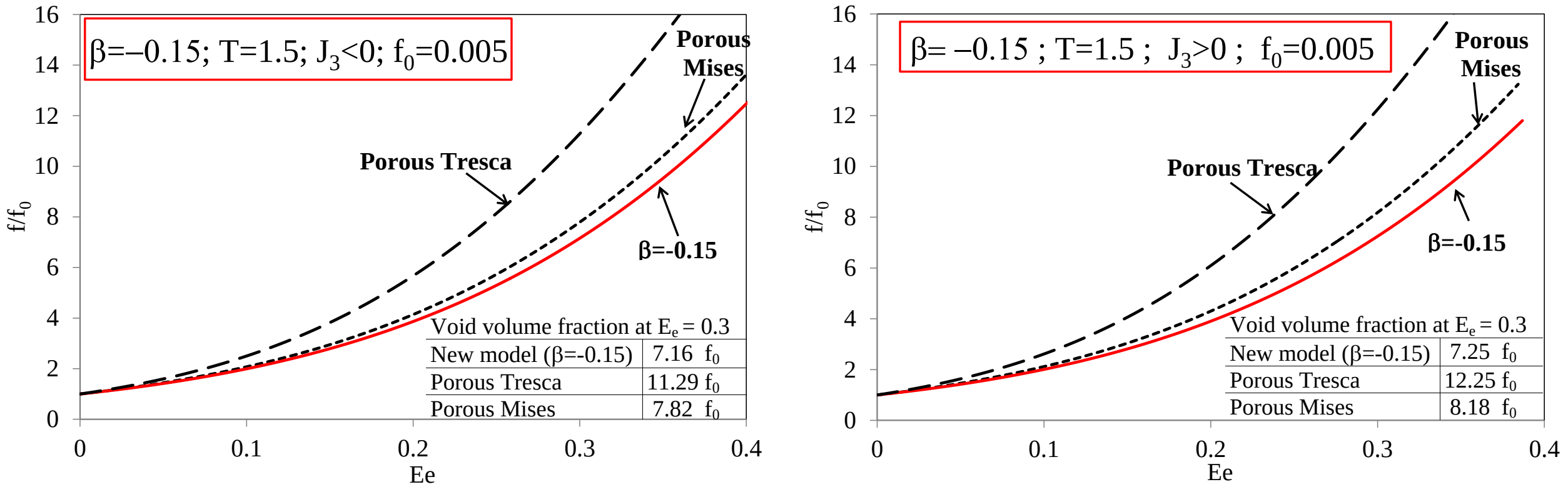

(a) (b)

**Fig.11**. Comparison between the evolution of the void volume fraction with equivalent plastic strain Ee (void growth) for axisymmetric stress states and fixed stress triaxiality T= 1.5 for a porous von Mises material (according to Cazacu et al. (2013)), a porous Tresca material (using Cazacu et al. (2014)) new criterion for porous solids (Eq. (28)-(33)) for a material with matrix characterized by β= -0.15 for (a) axisymmetric loadings such that $J_3^\Sigma \le 0$ ($\Sigma_{11} = \Sigma_{22} \ge \Sigma_{33}$) and (b) axisymmetric loadings such that $J_3^\Sigma \ge 0$ ($\Sigma_{11} = \Sigma_{22} \le \Sigma_{33}$). Initial porosity is the same in all materials: $f_0$=0.005. For the porous material with β= -0.15 there is practically no influence of $J_3^\Sigma$.

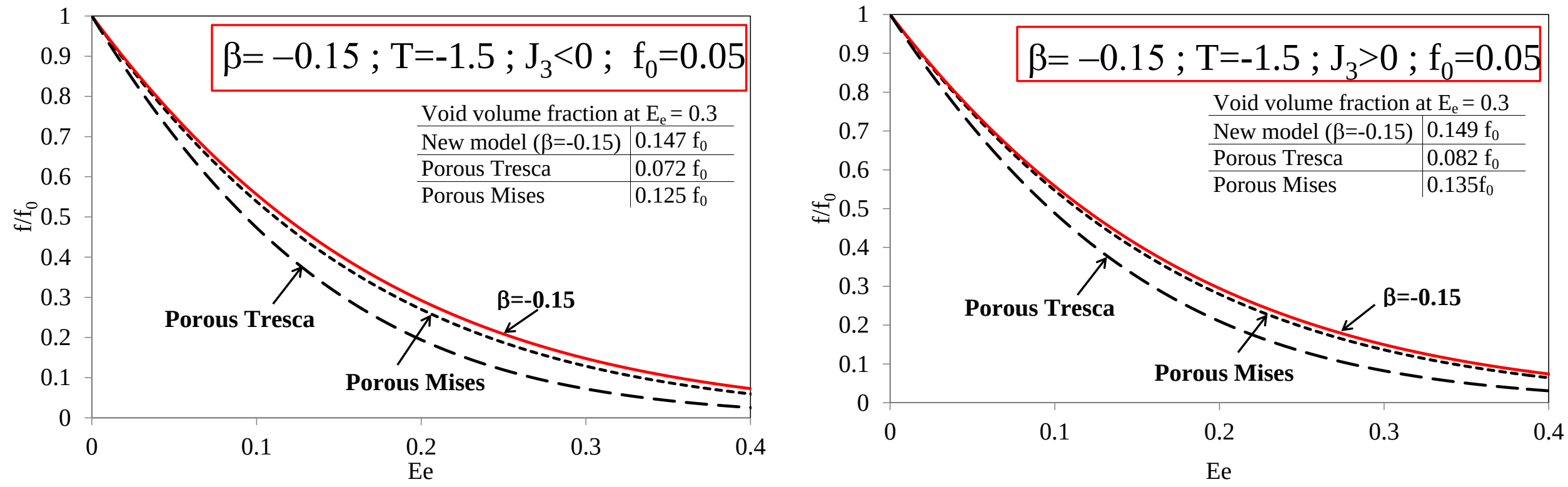

(a) (b)

**Fig. 12**. Comparison between the evolution of the void volume fraction with equivalent plastic strain Ee (void collapse) for axisymmetric stress states and fixed stress triaxiality T= -1.5 for a porous von Mises material (according to Cazacu et al. (2013)), a porous Tresca material (using Cazacu et al. (2014)) new criterion for porous solids (Eq. (28)-(33)) for a material with matrix characterized by β= -0.15 for (a) axisymmetric loadings such that $J_3^\Sigma \le 0$ ($\Sigma_{11} = \Sigma_{22} \ge \Sigma_{33}$) and (b) axisymmetric loadings such that $J_3^\Sigma \ge 0$ ($\Sigma_{11} = \Sigma_{22} \le \Sigma_{33}$). Initial porosity is the same in all materials: $f_0$=0.005. For the porous material with β= -0.15 there is practically no influence of $J_3^\Sigma$.

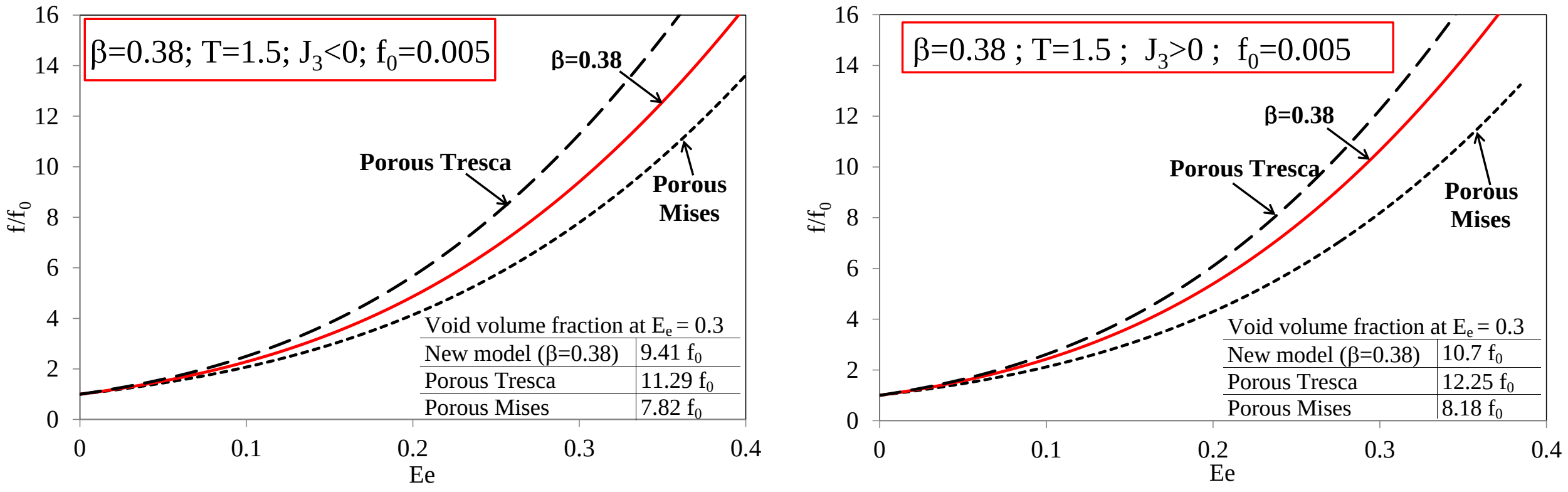

(a) (b)

**Fig. 13**. Comparison between the evolution of the void volume fraction with equivalent plastic strain Ee for axisymmetric stress states and fixed stress triaxiality T=1.5 predicted by the porous Mises model (Cazacu et al. (2013)), the porous Tresca criterion (Cazacu et al. (2014)) and the new criterion for porous solids (Eq. (28)-(33)) for a material with matrix characterized by β=0.38 for: (a) axisymmetric loadings such that $J_3^\Sigma \leq 0$ ($\Sigma_{11} = \Sigma_{22} \geq \Sigma_{33}$) and (b) axisymmetric loadings such that $J_3^\Sigma \geq 0$ ($\Sigma_{11} = \Sigma_{22} \leq \Sigma_{33}$). Initial porosity the same in all materials: $f_0$=0.005.

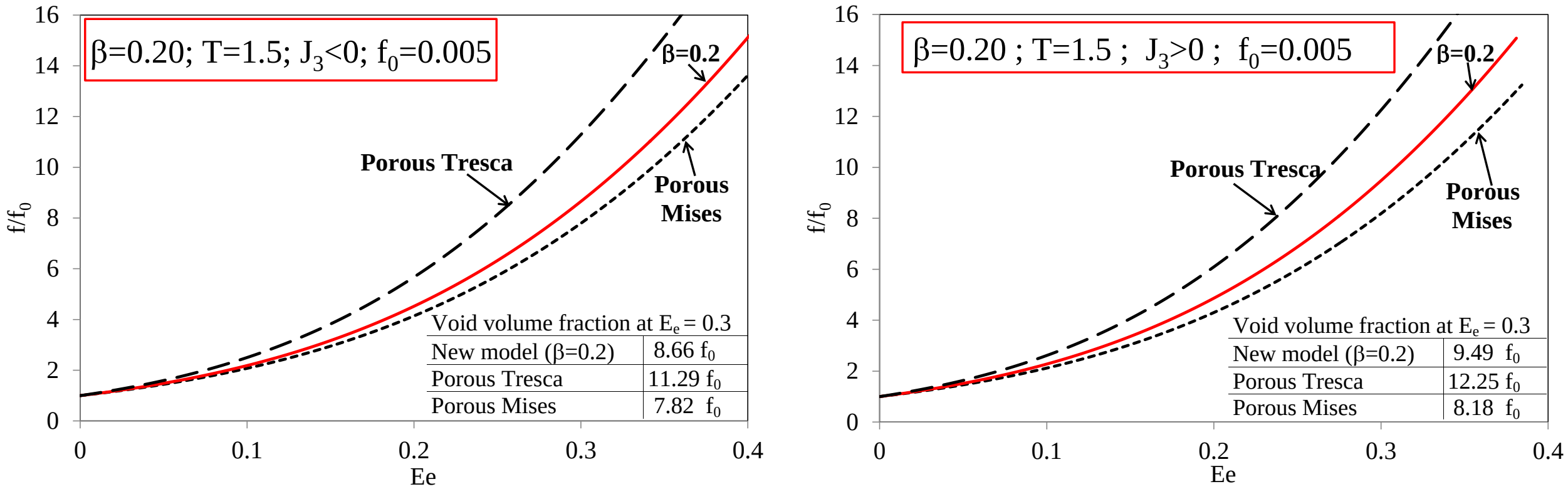

(a) (b)

**Fig. 14**. Comparison between the evolution of the void volume fraction with equivalent plastic strain Ee for axisymmetric stress states and fixed stress triaxiality T=1.5 for a porous von Mises material (according to Cazacu et al. (2013)), a porous Tresca material (using Cazacu et al. (2014)) new criterion for porous solids (Eq. (28)-(33)) for a material with matrix characterized by β=0.20 for (a) axisymmetric loadings such that $J_3^\Sigma \leq 0$ ($\Sigma_{11} = \Sigma_{22} \geq \Sigma_{33}$) and (b) axisymmetric loadings such that $J_3^\Sigma \geq 0$ ($\Sigma_{11} = \Sigma_{22} \leq \Sigma_{33}$). Initial porosity is the same in all materials: $f_0$=0.005.

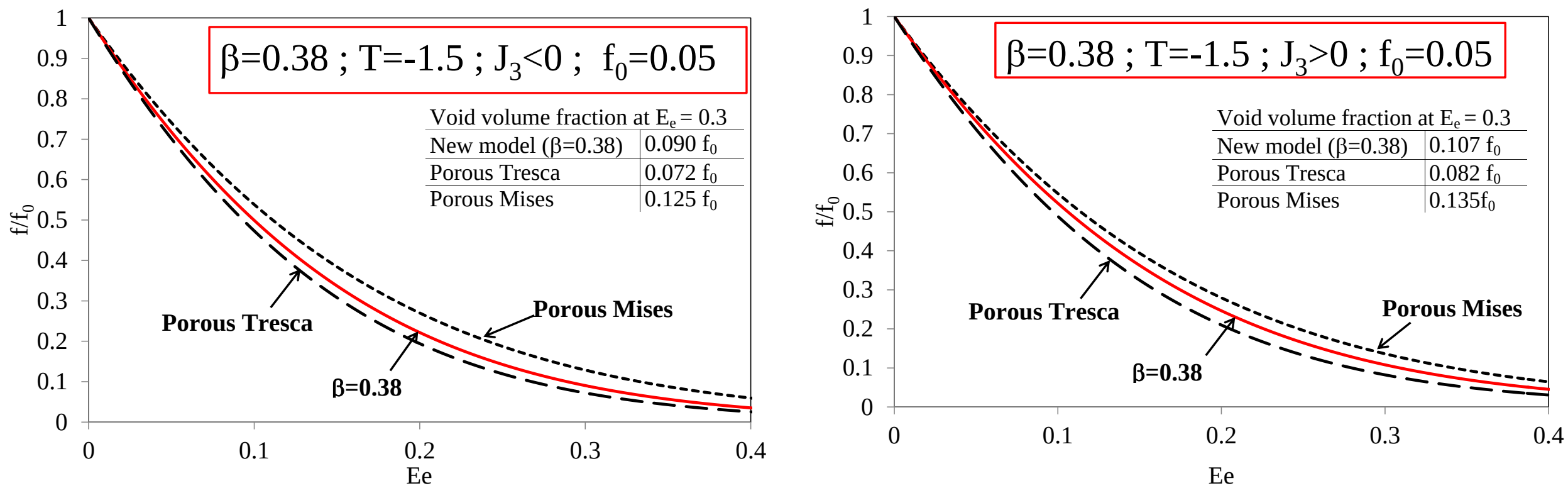

(a) (b)

**Fig. 15**. Comparison between the evolution of the void volume fraction with equivalent plastic strain Ee (void collapse) for axisymmetric stress states and fixed negative stress triaxiality T= -1.5 for a porous von Mises material (according to Cazacu et al. (2013)), a porous Tresca material (using Cazacu et al. (2014)) new criterion for porous solids (Eq. (28)-(33)) for a material with matrix characterized by β=0.38 for (a) axisymmetric loadings such that $J_3^{\Sigma} \leq 0$ $(\Sigma_{11} = \Sigma_{22} \geq \Sigma_{33})$ and (b) axisymmetric loadings such that $J_3^{\Sigma} \geq 0$ $(\Sigma_{11} = \Sigma_{22} \leq \Sigma_{33})$. Initial porosity is the same in all materials: $f_0$=0.005.

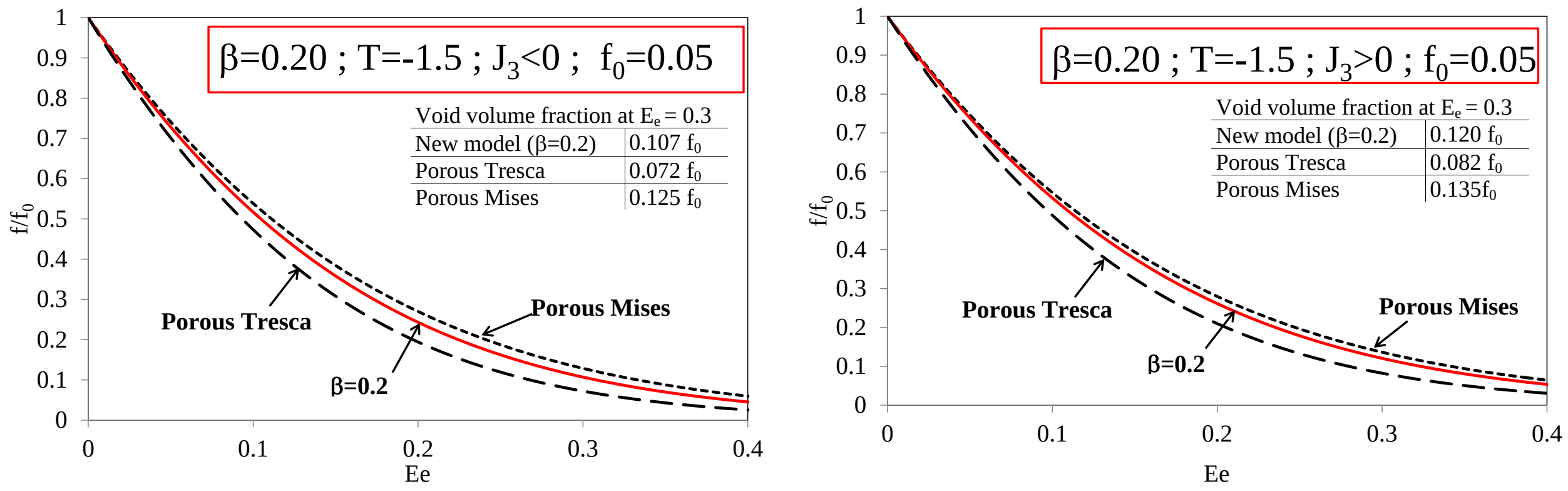

(a) (b)

**Fig. 16.** Comparison between the evolution of the void volume fraction with equivalent plastic strain Ee (void collapse) for axisymmetric stress states and fixed stress triaxiality T= -1.5 for a porous von Mises material (according to Cazacu et al. (2013)), a porous Tresca material (using Cazacu et al. (2014)) new criterion for porous solids (Eq. (28)-(33)) for a material with matrix characterized by β=0.2 for (a) axisymmetric loadings such that $J_3^{\Sigma} \leq 0$ $(\Sigma_{11} = \Sigma_{22} \geq \Sigma_{33})$ and (b) axisymmetric loadings such that $J_3^{\Sigma} \geq 0$ $(\Sigma_{11} = \Sigma_{22} \leq \Sigma_{33})$. Initial porosity is the same in all materials: $f_0$=0.005.

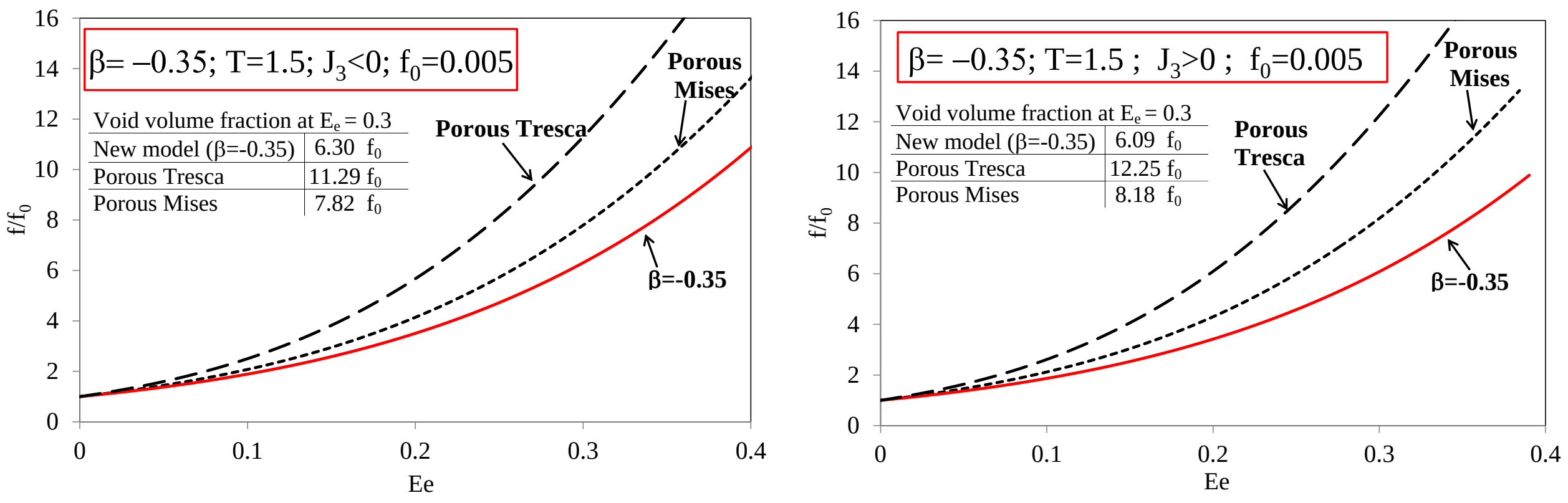

(a) (b)

**Fig. 17**. Comparison between the evolution of the void volume fraction with equivalent plastic strain Ee (void growth) for axisymmetric stress states and fixed stress triaxiality T= 1.5 for a porous von Mises material (according to Cazacu et al. (2013)), a porous Tresca material (using Cazacu et al. (2014)) new criterion for porous solids (Eq. (28)-(33)) for a material with matrix characterized by β= -0.35 for (a) axisymmetric loadings such that $J_3^\Sigma \leq 0$ $(\Sigma_{11} = \Sigma_{22} \geq \Sigma_{33})$ and (b) axisymmetric loadings such that $J_3^\Sigma \geq 0$ $(\Sigma_{11} = \Sigma_{22} \leq \Sigma_{33})$. Initial porosity is the same in all materials: $f_0$=0.005.

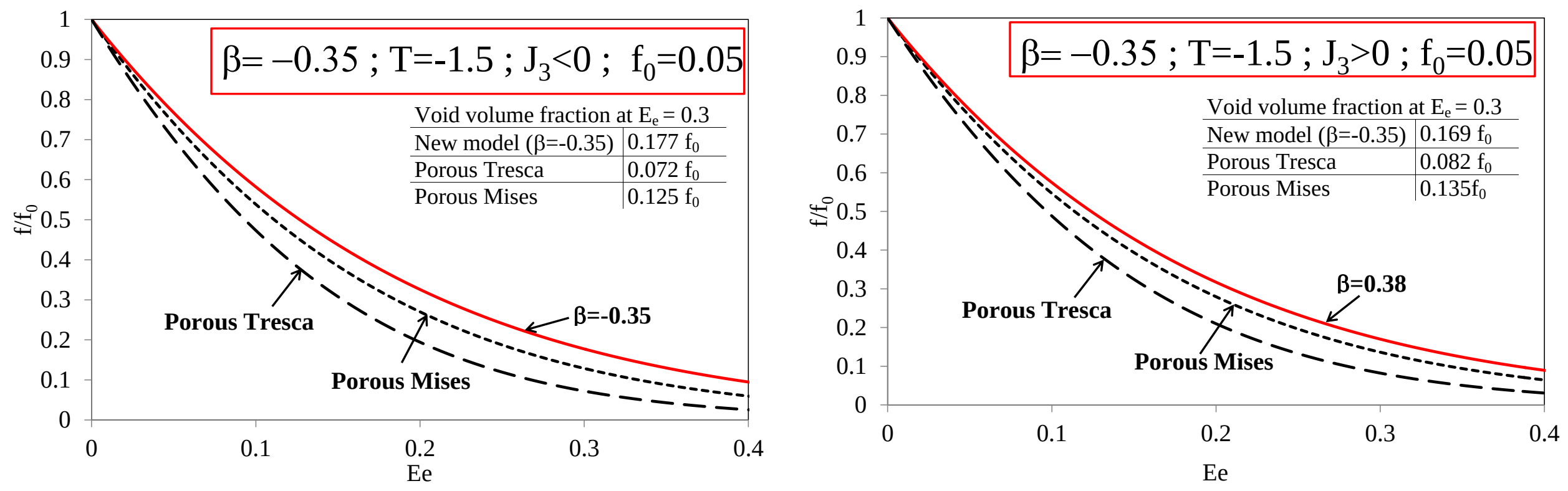

(a) (b)

**Fig. 18.** Comparison between the evolution of the void volume fraction with equivalent plastic strain Ee (void collapse) for axisymmetric stress states and fixed stress triaxiality T= -1.5 for a porous von Mises material (according to Cazacu et al. (2013)), a porous Tresca material (using Cazacu et al. (2014)) new criterion for porous solids (Eq. (28)-(33)) for a material with matrix characterized by β= -0.35 for (a) axisymmetric loadings such that $J_3^\Sigma \leq 0$ $(\Sigma_{11} = \Sigma_{22} \geq \Sigma_{33})$ and (b) axisymmetric loadings such that $J_3^\Sigma \geq 0$ $(\Sigma_{11} = \Sigma_{22} \leq \Sigma_{33})$. Initial porosity is the same in all materials: $f_0$=0.005.